\documentclass[journal=jctcce,manuscript=article]{achemso}

\usepackage[version=3]{mhchem} 
\usepackage{amsmath}
\usepackage{amssymb}
\usepackage{braket}

\newcommand{\rrangle}{\rangle\!\rangle}
\newcommand{\llangle}{\langle\!\langle}
\newcommand{\kket}[1]{\vert #1 \rrangle}
\newcommand{\bbra}[1]{\llangle #1 \vert}
\newcommand{\bbrakket}[1]{\llangle #1 \rrangle}
\newcommand{\sket}[1]{\vert #1 ]}
\newcommand{\sbra}[1]{[ #1 \vert}
\newcommand{\sbraket}[1]{[ #1 \rangle}
\newcommand{\brasket}[1]{\langle #1 ]}
\newcommand{\ee}{\text{e}}
\newcommand{\ii}{\text{i}}
\newcommand{\dd}{\text{d}}
\newcommand{\au}{\text{a.u.}}
\renewcommand{\Re}{\text{Re}}

\author{Thomas Bondo Pedersen}
\email{t.b.pedersen@kjemi.uio.no}
\author{Håkon Emil Kristiansen}
\author{Tilmann Bodenstein}
\author{Simen Kvaal}
\affiliation[Hylleraas Centre]
{Hylleraas Centre for Quantum Molecular Sciences, Department of Chemistry, University of Oslo, N-0315 Oslo, Norway}
\author{Øyvind Sigmundson Sch{\o}yen}
\affiliation{Department of Physics, University of Oslo, N-0315 Oslo, Norway}

\title[Interpretation of CC Dynamics]{Interpretation of Coupled-Cluster Many-Electron Dynamics
in Terms of Stationary States}

\begin{document}


\begin{abstract}
We demonstrate theoretically and numerically
that laser-driven many-electron dynamics, as described by
bivariational time-dependent coupled-cluster theory,
may be analyzed in terms of stationary-state populations.
Projectors heuristically defined from linear response theory and equation-of-motion coupled-cluster
theory are proposed for the calculation of stationary-state populations during
interaction with laser pulses or other external forces, and
conservation laws of the populations are discussed.
Numerical tests of the proposed projectors, involving both linear and nonlinear optical processes
for the He and Be atoms, and for the LiH, CH$^+$,
and LiF molecules, show that the laser-driven evolution of the stationary-state populations
at the coupled-cluster singles-and-doubles (CCSD) level is very close to that obtained by full
configuration-interaction theory
provided all stationary  states actively participating in the dynamics are sufficiently
well approximated. When double-excited states are important for the dynamics, the quality of the
CCSD results deteriorate. Observing that populations computed from the linear-response projector may
show spurious small-amplitude, high-frequency oscillations, the equation-of-motion 
projector emerges as the most promising approach to stationary-state populations.
\end{abstract}

\section{Introduction}

Providing unique time-resolved insights into electronic quantum dynamics,
and with the exciting prospect of detailed manipulation and control of
chemical reactions~\cite{Lepine2014},
increasing experimental and theoretical research efforts have been directed toward attosecond
science in the past couple of decades---see, e.g., Ref.~\citenum{Li2020} for a recent perspective.
While the initial step usually involves ionization induced by extreme-ultraviolet or near-infrared
laser pulses, \citeauthor{Hassan2016}~\cite{Hassan2016} have demonstrated that optical attosecond
pulses may be used to observe and control the dynamics of bound electrons with little
or no ionization probability. Whether ionization plays a role or not, the rapid
development of experimental methodology creates a strong demand for
explicitly time-dependent
quantum chemical methods that can accurately simulate the ultrafast many-electron
dynamics driven by ultrashort laser pulses.

While real-time time-dependent density-functional theory (TDDFT)~\cite{Runge1984,VanLeeuwen1999,Ullrich2012}
is a highly attractive option from the viewpoint of computational efficiency,
it suffers from a number of deficiencies caused largely by the reliance on the adiabatic
approximation in most practical applications~\cite{Li2020b}.
Improved accuracy can be achieved
with wave function-based methods at the expense of increased computational costs~\cite{Li2020b}.
In a finite basis, the exact solution to the time-dependent electronic Schr{\"o}dinger equation
is the full configuration interaction (FCI) wave function whose computational
complexity, unfortunately, increases exponentially with the number of electrons.
We are thus forced to introduce approximations.
The perhaps most widely used time-dependent wave function approximation for simulating many-electron dynamics
is multiconfigurational time-dependent Hartree-Fock (MCTDHF)
theory~\cite{Zanghellini2003,Kato2004,Meyer2009,Hochstuhl2014} and the related
time-dependent complete active space self-consistent field
(TD-CASSCF) and restricted active space (TD-RASSCF)
methods~\cite{Hochstuhl2014,Sato2013,Miyagi2013,Miyagi2014}.
Restricting the participating Slater determinants to those
that can be generated from a fixed number of electrons and a carefully selected
active space of (time-dependent) spin orbitals, these methods still have the
FCI wave function at the heart, eventually facing the exponential scaling wall
as the number of active electrons and orbitals is increased.

Coupled-cluster (CC) theory offers a more gentle polynomial-scaling
hierarchy of approximations that converge to the FCI wave function.
Besides the differences in computational complexity, the two methods differ
in the sense that MCTDHF captures static (strong) correlation, wheras single-reference
CC theory aims at dynamical correlation effects.
Yielding energies, structures, and properties with excellent accuracy
for both ground- and excited states of weakly correlated systems,
it has become one of the most trusted methods of molecular quantum chemistry~\cite{Bartlett2007}.
Recent years have witnessed increasing interest in time-dependent CC (TDCC)
theory~\cite{Hoodbhoy1978,Hoodbhoy1979,Schonhammer1978,Dalgaard1983,Arponen1983} 
for numerical simulations of many-body quantum dynamics in
nuclear~\cite{Pigg2012}, and
atomic
and
molecular~\cite{Huber2011,Kvaal2012,Nascimento2016,Nascimento2017,Nascimento2019,Koulias2019,Park2019,Pedersen2019,Kristiansen2020,Sato2018,Pathak2020,Pathak2020a,Pathak2020b,Hansen2019,Hansen2020,Rehr2020,Skeidsvoll2020}
systems.
In addition, TDCC theory plays a key role in recent work on
finite-temperature CC theory for molecular~\cite{White2019,White2020}
and extended~\cite{Hummel2018}
systems.
While the papers by Christiansen and coworkers~\cite{Hansen2019,Hansen2020}
are concerned with vibrational CC theory and that of \citeauthor{Pigg2012}~\cite{Pigg2012}
with nucleon dynamics,
the remaining papers are focused on the dynamics of atomic and molecular electrons exposed to
electromagnetic fields such as ultrashort laser pulses.

In many cases, the main goal is to compute absorption (or emission)
spectra~\cite{Nascimento2016,Nascimento2017,Nascimento2019,Koulias2019,Park2019,Rehr2020,Skeidsvoll2020}
by Fourier transformation of the induced dipole moment. This requires the calculation
of the induced dipole moment for extended periods of time \emph{after} the perturbing
field or laser pulse has been turned off. While decisive for the features observed in the final spectrum,
the dynamics \emph{during} the interaction with the laser pulse is rarely analyzed in detail.
Processes that occur during the pulse, such as high harmonic generation and ionization,
are studied using TDCC theory in Refs.~\citenum{Sato2018,Pathak2020,Pathak2020a,Pathak2020b}.
Since energy is the physical quantity associated with time translations,
textbook analyses of such interactions are naturally performed in terms of the population of
the energy eigenstates---the stationary states---of the field-free particle system, see, e.g.,
Ref.~\citenum{Loudon2000}.
However, many-body theories such as the TDFCI, MCTDHF, and TDCC theories
do not express the wave function as a superposition of stationary states, making the
analysis difficult to perform in simulations. Moreover, when approximations are introduced
(truncation of the many-body expansion), the stationary states are hard to define
precisely for nonlinear parameterizations such as MCTDHF and TDCC theory.
The problem is particularly pronounced for approximate methods where the orbitals
are time-dependent such that a different subspace of the full configuration space
is spanned in each time step of a simulation, leading to energies and eigenvectors of
the Hamilton matrix that vary depending on the laser pulse applied to the system~\cite{Padmanaban2008}.
This implies, for example, that identification of stationary-state energies by
Fourier transformation of the post-pulse autocorrelation function leads to pulse-dependent results.
Still, several reports of population transfer
during interaction with laser pulses have been reported recently~\cite{Li2014,Haxton2014,Greenman2017}
using MCTDHF theory.

The natural approach would be to define the stationary states from the zero-field
Hamiltonian and zero-field wave function using, e.g., linear response theory~\cite{Olsen1985} or 
orthogonality-constrained imaginary
time propagation~\cite{Caillat2005}.
The latter approach was investigated recently within the framework of MCTDHF theory
by \citeauthor{Lotstedt2020}~\cite{Lotstedt2020}, who found that the stationary-state
populations oscillate even after the pulse is turned off unless a sufficiently large
number of active orbitals is included in the wave function expansion.
In this work, we use both CC linear response theory~\cite{Dalgaard1983,Koch1990,Christiansen1998}
and equation-of-motion CC (EOMCC) theory~\cite{Emrich1981,Stanton1993,Krylov2008,Izsak2019}
to propose projectors whose expectation values yield stationary-state populations.
Test simulations are presented with different laser pulses and the TDCC results are 
compared with the exact (TDFCI) results.

The paper is organized as follows. In Section \ref{sec:theory}, we briefly outline
exact quantum dynamics in the basis of energy eigenstates and use analogies to
propose projectors whose expectation values can be interpreted as
stationary-state populations within TDCC theory.
Technical details of the numerical simulations are given in Section \ref{sec:computational_details}
and numerical results are presented and discussed in Section \ref{sec:results_and_discussion}
for atoms and diatomic molecules in few-cycles laser pulses, including chirped pulses.
Concluding remarks are given in Section \ref{sec:concluding_remarks}.

\section{Theory}
\label{sec:theory}

\subsection{Recapitulation of exact quantum dynamics}
\label{subsec:QD_recap}

Laser-driven quantum dynamics of a particle system is usually interpreted in terms of 
stationary states $\ket{n}$ defined as solutions of the time-independent
Schr{\"o}dinger equation
\begin{equation}
    H_0 \ket{n} = E_n \ket{n}
\end{equation}
where $H_0$ is the time-independent Hamiltonian of the particle system and $E_n$ is
the energy of the stationary state $\ket{n}$. The stationary states evolve in time
according to
\begin{equation}
    \ket{n(t)} = \ket{n}\ee^{-\ii E_nt}
\end{equation}
and are assumed to
form a complete orthonormal set such that
\begin{equation}
   P_n = \ket{n}\!\bra{n} = \ket{n(t)}\!\bra{n(t)}, \qquad \sum_n P_n = 1
\end{equation}
where $1$ is the identity operator. Note that the continuum is formally included
in the summation over states.

The time evolution of the particle
system is determined by the time-dependent Schr{\"o}dinger equation
(using atomic units ($\au$) throughout),
\begin{equation}
    H(t) \ket{\Psi(t)} = \ii \ket{\dot{\Psi}(t)}, \qquad \ket{\Psi(t=0)} = \ket{\Psi_0}
\end{equation}
where $\ket{\Psi(t)}$ is the normalized state of the system with known initial value 
$\ket{\Psi_0}$, and the dot denotes the time derivative.
Within semi-classical radiation theory, the time-dependence of the Hamiltonian
\begin{equation}
   H(t) = H_0 + V(t)
\end{equation}
stems from the operator $V(t)$ describing the interaction between the particle system
and external electromagnetic fields.

Expressing the time-dependent state as a superposition of stationary states,
\begin{equation}
    \ket{\Psi(t)} = \sum_n P_n \ket{\Psi(t)} = \sum_n \ket{n} C_n(t) \ee^{-\ii E_nt}
\end{equation}
where $C_n(t) = \braket{n(t) \vert \Psi(t)}$,
the time-dependent Scr{\"o}dinger equation may be recast as
an ordinary differential equation
\begin{equation}
\label{eq:tdse_model}
   \ii \dot{C}_n(t) = \sum_m \braket{n \vert V(t) \vert m} C_m(t) \ee^{\ii (E_n - E_m)t}
\end{equation}
with the initial conditions $C_n(0) = \braket{n \vert \Psi_0}$.

The population of stationary state $\ket{n}$ at any time $t \geq 0$ may be determined
as the expectation value of the projection operator $P_n$,
\begin{equation}
    p_n(t) = \braket{\Psi(t) \vert P_n \vert \Psi(t)}
    = \vert C_n(t) \vert^2
\end{equation}
which is real and non-negative.
Since the state is assumed normalized, $\braket{\Psi(t) \vert \Psi(t)} = 1$,
the stationary-state populations sum up to one, $\sum_n p_n(t) = 1$, and, hence,
are bounded from above as well as from below: $0 \leq p_n(t) \leq 1$.

It follows from the time-dependent Schr{\"o}dinger equation that the expectation
value of some operator, say $B$, evolves according to the Ehrenfest theorem
\begin{equation}
   \frac{\dd}{\dd t} \braket{\Psi(t) \vert B \vert \Psi(t)} =
   -\ii \braket{\Psi(t) \vert [B,H] \vert \Psi(t)} +
   \braket{\Psi(t) \vert \frac{\partial B}{\partial t} \vert \Psi(t)}
\end{equation}
where the last term vanishes when $B$ is a time-independent operator such as
a stationary-state projector. Hence, stationary-state populations are conserved
in the absence of external forces, as the projectors commute with the
time-independent Hamiltonian,
$[P_n,H_0] = 0$.

Stationary-state populations are required to identify transient phenomena
such as Rabi oscillations in simulations,
and may be used to determine the composition of the quantum state resulting
from the application of a short pump laser, facilitating interpretation of
spectra recorded by means of a subsequent probe laser. 
The stationary-state populations can be controlled by varying the pump laser parameters
such as peak intensity, shape, and duration. Predicting final
populations with varying laser parameters thus becomes a central computational
goal.

Unfortunately, computing all stationary states of a particle system followed by integration of the 
time-dependent Schr{\"o}dinger equation presents an insurmountable challenge, even if the number of
stationary states is kept finite through the use of a finite set of basis functions.
In practice, the quantum state is parameterized in a finite-dimensional Hilbert
space spanned by well-defined basis vectors (Slater determinants, in the case of electronic
systems) rather than stationary states. Populations of a few selected (low-lying) stationary states
can then be computed as a function of time with varying pump laser parameters.
This procedure is easily implemented for any electronic-structure method with
explicit parameterization of orthogonal ground- and excited-state wave functions
and has been used recently by, e.g., \citeauthor{Peng2018}~\cite{Peng2018} within
time-dependent configuration interaction theory to predict populations of stationary
electronic states of the rigid decacene molecule with varying laser parameters.
In the following, we will present an approach to the calculation of stationary-state
populations within the framework of TDCC theory.

\subsection{The time-dependent coupled-cluster state vector}

Our starting point is
Arponen's time-dependent bivariational formulation of CC
theory~\cite{Arponen1983} within the clamped-nucleus Born-Oppenheimer
approximation. This allows us to parameterize the CC ket and bra wave functions
as \emph{independent} approximations to the FCI
wave function and its hermitian conjugate.
The quantum state of an atomic or molecular many-electron system at time $t$
is then represented by the TDCC state
vector~\cite{Pedersen2019}
\begin{equation}
  \kket{S(t)} = \frac{1}{\sqrt{2}}
   \begin{pmatrix}
     \ket{\Psi(t)} \\
     \ket{\tilde{\Psi}(t)}
   \end{pmatrix}
\end{equation}
where the component functions are defined by
\begin{align}
    &\ket{\Psi(t)} = \ee^{T(t)}\ket{\Phi_0}\ee^{\tau_0(t)} \\
    &\bra{\tilde{\Psi}(t)} =
    \ee^{-\tau_0(t)} \bra{\Phi_0}(\lambda_0(t) + \Lambda(t))
    \ee^{-T(t)}
\end{align}
Although the reference determinant $\ket{\Phi_0}$ should be constructed from
time-dependent orthonormal~\cite{Hoodbhoy1978,Pedersen1999,Sato2018} or
biorthonormal~\cite{Pedersen2001,Kvaal2012,Kristiansen2020} orbitals to capture
the main effects of interactions between the electrons and external fields,
we shall in this work use the static Hartree-Fock (HF) ground-state
determinant for simplicity.
As long as the external field does not lead to nearly complete depletion
of the ground state, we found in Ref.~\citenum{Kristiansen2020} that the results obtained with
static and dynamic orbitals are virtually identical.
In addition, using static HF reference orbitals allows us to exploit well-known CC theories for
excited states, as discussed in more detail below, and we avoid the complexity
of computing overlaps between determinants in different non-orthogonal orbital bases.

The cluster operators are defined as
\begin{equation}
   T(t) = \sum_\mu \tau_\mu(t) X_\mu, \qquad
   \Lambda(t) = \sum_\mu \lambda_\mu Y_\mu^\dagger
\end{equation}
where $\mu > 0$ labels excitations out of the reference determinant,
\begin{equation}
   \ket{\Phi_\mu} = X_\mu \ket{\Phi_0}, \qquad
   \bra{\tilde{\Phi}_\mu} = \bra{\Phi_0} Y_\mu^\dagger, \qquad
   \braket{\tilde{\Phi}_\mu \vert \Phi_\nu} = \delta_{\mu\nu}
\end{equation}
If the cluster operators include all excited determinants, the CC state
becomes equivalent to the exact wave function, the full configuration
interaction (FCI) wave function. Approximations are obtained by
truncating the cluster operators after singles to give the CCS method,
after singles and doubles to give the CCSD model, and so on.
Since the HF reference determinant is static, the time-dependence of the
cluster operators is carried by the amplitudes $\tau_\mu(t)$ and
$\lambda_\mu(t)$ only.
The amplitude $\tau_0(t)$ is a phase parameter related to the
socalled
quasienergy~\cite{Sambe1973,Sasagane1993} and $\lambda_0(t)$ determines
the normalization of the state, as dicussed in more detail below.

In the closed-shell spin-restricted CCSD model, the singles and 
doubles excitation and deexcitation
operators are defined as~\cite{MEST}
\begin{alignat}{2}
   &X_{ai} = E_{ai}, &\qquad &X_{aibj} = E_{ai}E_{bj} \\
   &Y_{ai}^\dagger = \frac{1}{2}X_{ai}^\dagger, &\qquad
   &Y_{aibj}^\dagger = \frac{1}{6(1 + \delta_{ab}\delta_{ij})}\left(
                   2X_{aibj}^\dagger + X_{ajbi}^\dagger
               \right)
\end{alignat}
where $i$, $j$ and $a$, $b$ refer to occupied and virtual spatial HF orbitals,
respectively, and
\begin{equation}
   E_{ai} = c^\dagger_{a\alpha}c_{i\alpha} + c^\dagger_{a\beta}c_{i\beta}
\end{equation}
is a unitary group generator expressed in terms of the elementary second-quantization
spin-orbital ($\alpha$ and $\beta$ here refer to the spin-up and spin-down states)
creation and annihilation operators.

The equations of motion for the amplitudes are derived from the time-dependent
bivariational principle and are given by~\cite{Arponen1983,Pedersen2019}
\begin{alignat}{2}
\label{eq:eom_tau}
  &\ii\dot{\tau}_0
  = \braket{\Phi_0 \vert \ee^{-T(t)}H(t)\ee^{T(t)}\vert \Phi_0},
  &\qquad
  &\ii\dot{\tau}_\mu
  = \braket{\tilde{\Phi}_\mu \vert \ee^{-T(t)}H(t)\ee^{T(t)} \vert \Phi_0}
  \\
\label{eq:eom_lambda}
 -&\ii\dot{\lambda}_0 = 0,
  &\qquad
 -&\ii\dot{\lambda}_\mu
  = \braket{\tilde{\Psi}(t) \vert [H(t),X_\mu] \vert \Psi(t)}
\end{alignat}
where the dot denotes the time derivative, and
\begin{equation}
   H(t) = H_0 + V(t)
\end{equation}
While $H_0$ is the time-independent molecular electronic Hamiltonian in the
clamped-nucleus Born-Oppenheimer approximation, $V(t)$ describes
the interaction of the electrons with explicitly time-dependent external fields
in the semi-classical approximation.
Note that the normalization amplitude $\lambda_0$ is constant.

The equations of motion \eqref{eq:eom_tau} and \eqref{eq:eom_lambda} must be
integrated with suitable initial conditions. In this work, we use the CC
ground state
\begin{equation}
   \kket{S(t=0)} = \kket{S_0} = \frac{1}{\sqrt{2}}
   \begin{pmatrix}
       \ket{\Psi_0} \\ \ket{\tilde{\Psi}_0}
   \end{pmatrix}
\end{equation}
where
\begin{alignat}{2}
   &\ket{\Psi_0} = \ee^{T_0} \ket{\Phi_0}, &\qquad
   &T_0 = \sum_\mu \tau_\mu^0 X_\mu \\
   &\bra{\tilde{\Psi}_0} = \bra{\Phi_0} (\lambda_0 + \Lambda_0) \ee^{-T_0},
        &\qquad
   &\Lambda_0 = \sum_\mu \lambda_\mu^0 Y_\mu^\dagger
\end{alignat}
The ground-state amplitudes satisfy the stationary CC equations
\begin{align}
   &0 = \braket{\tilde{\Phi}_\mu \vert \ee^{-T_0}H_0\ee^{T_0}\vert \Phi_0} \\
   &0 = \braket{\tilde{\Psi}_0 \vert [H_0,X_\mu] \vert \Psi_0}
\end{align}
and $\tau_0(t=0) = 0$ such that, in the absence of external perturbations,
the time-dependence of the TDCC state vector correctly becomes
$\kket{S(t)} = \kket{S_0}\exp(-\ii E_0t)$ where $E_0$ is the CC ground-state energy
\begin{equation}
   E_0 = \braket{\Phi_0 \vert \ee^{-T_0}H_0\ee^{T_0}\vert \Phi_0}
\end{equation}

\subsection{Interpretation}

We now introduce the indefinite inner
product~\cite{Pedersen2019}
\begin{equation}
\label{eq:iip}
   \bbrakket{S_1 \vert S_2} =
   \frac{1}{2}
   \begin{pmatrix}
      \bra{\tilde{\Psi}_1} & \bra{\Psi_1}
   \end{pmatrix}
   \begin{pmatrix}
      \ket{\Psi_2} \\ \ket{\tilde{\Psi}_2}
   \end{pmatrix}
   =
   \frac{1}{2} \braket{\tilde{\Psi}_1 \vert \Psi_2} +
   \frac{1}{2} \braket{\tilde{\Psi}_2 \vert \Psi_1}^*
\end{equation}
with respect to which the TDCC state vector is normalized
\begin{equation}
   \bbrakket{S(t) \vert S(t)} = 1
\end{equation}
provided we choose $\Re (\lambda_0) = 1$. In practice, we choose $\lambda_0 = 1$.

The indefinite inner product induces the expectation value
expression~\cite{Pedersen2019}
\begin{equation}
\label{eq:expval_cc}
   \braket{C}(t) = \bbrakket{S(t) \vert \hat{C} \vert S(t)}
   = \frac{1}{2} \braket{\tilde{\Psi}(t) \vert C \vert \Psi(t)}
   + \frac{1}{2} \braket{\tilde{\Psi}(t) \vert C^\dagger \vert \Psi(t)}^*
\end{equation}
where the two-component form of the quantum-mechanical operator $C$ reads
\begin{equation}
\label{eq:property_operator_2c}
   \hat{C} = \begin{pmatrix} C & 0 \\ 0 & C \end{pmatrix}
\end{equation}
While  the expectation value of an anti-hermitian operator $C^\dagger = -C$ is
imaginary,
the expectation value of a hermitian operator $C^\dagger = C$ is real.
This symmetrized form of the CC expectation value was first introduced
by Pedersen and Koch~\cite{Pedersen1997} in order to ensure
correct symmetries, including time-reversal symmetry, of CC response functions.
Using the expectation value expression to compute the electric dipole moment
induced by an external laser field, absorption spectra can be obtained by
Fourier transformation.

Given a set of orthonormal excited-state vectors $\kket{E_n}$, which
are orthogonal to the ground state with respect to the indefinite
inner product, we may define the projection operator
\begin{equation}
\label{eq:P_n}
   \hat{P}_n = \kket{E_n}\bbra{E_n}
\end{equation}
and compute the population $p_n(t)$ of excited state $n$ at time $t$ as the 
expectation value
\begin{equation}
\label{eq:excited_pop}
   p_n(t) = \bbrakket{S(t) \vert \hat{P}_n \vert S(t)}
\end{equation}
This would provide a time-resolved picture of the populations of excited
states within TDCC theory.
Unfortunately, a fully consistent set of CC excited-state vectors is not
known. 

There are two distinct approaches to excited states in common use 
within CC theory today.~\cite{Sneskov2012}
One is the 
EOMCC~\cite{Emrich1981,Stanton1993,Krylov2008,Izsak2019} approach
where the excited states are \emph{parameterized} explicitly in terms
of linear excitation and de-excitation operators, which generate the
excited-state vectors from the ground state.
While making it straightforward to express the projection operator in 
eq.~\eqref{eq:P_n}, 
the linear \emph{Ansatz} of EOMCC theory leads to size-intensivity issues
in transition moments and response properties such as
polarizabilities.~\cite{Kobayashi1994,Koch1994,Nanda2018}

The other approach is CC response
theory~\cite{Dalgaard1983,Koch1990,Christiansen1998}
where the time-dependent Schr{\"o}dinger equation is solved
in the frequency domain using a combination of Fourier transformation and
adiabatic perturbation theory. The amplitudes at each order are
then used to express linear, quadratic, and higher-order response functions.
This leads to the identification of excitation energies
and one- and multiphoton transition moments by analogy with response theory
for exact stationary states.~\cite{Olsen1985}
It does not, however,
lead to explicit expressions for the excited-state vectors, making it
impossible to construct projection operators of the form \eqref{eq:P_n}.
While the excitation energies from EOMCC theory and CC response theory are
identical and properly size-intensive,
the transition moments differ and no size-extensivity issues
are present in CC response theory.~\cite{Kobayashi1994,Koch1994}

We note in passing that \citeauthor{Hansen2020}~\cite{Hansen2020}
offer a detailed discussion
of size-extensivity (and, by extension, size-intensivity) and size-consistency
issues within CC theory
using the more concise and accessible concepts of additive and multiplicative separability.

\subsection{Projection operators from EOMCC theory}

Exploiting that EOMCC theory provides an explicit parameterization of
``left'' (bra) and ``right'' (ket) excited states,
we investigate the EOMCC projector defined as
\begin{equation}
\label{eq:conventional_eom_projector}
   \hat{P}_n =
   \begin{pmatrix}
      \ket{\Psi_n}\bra{\tilde{\Psi}_n} & 0 \\
      0 & \ket{\tilde{\Psi}_n}\bra{\Psi_n}
   \end{pmatrix}
\end{equation}
where
\begin{equation}
\label{eq:eom_rl_states}
   \ket{\Psi_n} = \ket{\Psi_n^x} + {\cal R}^n_0 \ket{\Psi_0}
   = \left(R_n + {\cal R}^n_0\right) \ket{\Psi_0}, \qquad
   \bra{\tilde{\Psi}_n} = \bra{\Phi_0} L_n \ee^{-T_0}
\end{equation}
and the linear excitation and de-excitation operators are given by
\begin{equation}
   R_n = \sum_\mu {\cal R}^n_\mu X_\mu, \qquad
   L_n = \sum_\mu {\cal L}^n_\mu Y_\mu^\dagger
\end{equation}
The EOMCC excitation and de-excitation operators are truncated at the same
level as the underlying CC ground-state cluster operators $T_0$ and
$\Lambda_0$.
The ground-state component reads
\begin{equation}
   {\cal R}^n_0 = -\braket{\tilde{\Psi}_0\vert \Psi_n^x}
   = -\braket{\tilde{\Psi}_0 \vert R_n \vert \Psi_0}
\end{equation}
The EOMCC amplitudes are determined from the nonhermitian eigenvalue
problem
\begin{equation}
\label{eq:eigen}
   \boldsymbol{{\cal L}A{\cal R}} = \boldsymbol{\Delta E}, \qquad
   \boldsymbol{{\cal LR}} = \boldsymbol{1}
\end{equation}
where $\boldsymbol{1}$ is the unit matrix and $\boldsymbol{\Delta E}$ 
is a diagonal matrix with the excitation energies $\Delta E_n = E_n - E_0$
as elements. The elements of the nonhermitian Jacobian matrix are defined by
\begin{equation}
\label{eq:cc_jacobian}
   A_{\mu\nu} = \braket{\tilde{\Phi}_\mu \vert [\bar{H}_0, X_\nu] \vert \Phi_0}
\end{equation}
where $\bar{H}_0 = \exp(-T_0) H_0 \exp(T_0)$.

Note that the components of the EOMCC excited-state vector are
biorthonormal,
\begin{equation}
\label{eq:eom_orth}
   \braket{\tilde{\Psi}_n \vert \Psi_m} = \delta_{nm}
\end{equation}
and orthogonal to the CC ground state in the sense of
\begin{equation}
\label{eq:eom_orth_gs}
   \braket{\tilde{\Psi}_n \vert \Psi_0}  =
   \braket{\tilde{\Psi}_0 \vert \Psi_n} = 0
\end{equation}
This implies that the EOMCC projectors in eq.~\eqref{eq:conventional_eom_projector}
are hermitian with respect to the indefinite inner product,
annihilate the ground state, $\hat{P}_n \kket{S_0} = 0$,
and are idempotent and orthogonal
\begin{equation}
\label{eq:orthogonality}
   \hat{P}_n \hat{P}_m = \delta_{nm}\hat{P}_n
\end{equation}
In the limit of untruncated 
cluster operators, it is readily verified---using the orthonormality
of the left and right eigenvectors in eq.~\eqref{eq:eigen}---that
the EOMCC projectors satisfy the completeness relation
\begin{equation}
   \sum_n \hat{P}_n = 
   \begin{pmatrix}
      1 & 0 \\
      0 & 1
   \end{pmatrix}
   -
   \begin{pmatrix}
      \ket{\Psi_0}\bra{\tilde{\Psi}_0} & 0 \\
      0 & \ket{\tilde{\Psi}_0}\bra{\Psi_0}
   \end{pmatrix}
\end{equation}
If the cluster operators are truncated, the right-hand side must
be corrected for excited determinants beyond the CC truncation level
(e.g., triples, quadruples, etc.~for EOMCCSD).

The one-photon transition strength obtained from the EOMCC projector
is given by
\begin{equation}
   \bbrakket{S_0 \vert \hat{B} \hat{P}_n \hat{C} \vert S_0}
   =
   \frac{1}{2} \left(
   \braket{\tilde{\Psi}_0 \vert B \vert \Psi_n}
   \braket{\tilde{\Psi}_n \vert C \vert \Psi_0}
   +
   \braket{\tilde{\Psi}_0 \vert C \vert \Psi_n}^*
   \braket{\tilde{\Psi}_n \vert B \vert \Psi_0}^*
   \right)
\end{equation}
where $B$ and $C$ are hermitian operators representing electric or
magnetic multipole moments, and $\hat{B}$ and $\hat{C}$ are their
two-component forms defined in eq.~\eqref{eq:property_operator_2c}. 
The transition strength is properly symmetrized with respect to
simultaneous permutation of the multipole operators and complex conjugation,
$\bbrakket{S_0 \vert \hat{C} \hat{P}_n \hat{B} \vert S_0}^* =
\bbrakket{S_0 \vert \hat{B} \hat{P}_n \hat{C} \vert S_0}$,
and agrees with the commonly used expression in EOMCC theory,
which is based on a CI-like intepretation of the bra and ket states.
This expression yields the correct FCI limit but, as mentioned above,
the transition strength is \emph{not} properly size-intensive when the 
cluster operators are truncated~\cite{Kobayashi1994,Koch1994,Nanda2018}.

We may now use
eq.~\eqref{eq:conventional_eom_projector} to extract excited-state
populations from the TDCC state vector according to eq.~\eqref{eq:excited_pop}:
\begin{equation}
\label{eq:excited_pop_eom_conventional}
   p_n(t) = \Re\left(
       \braket{\tilde{\Psi}(t) \vert \Psi_n}
       \braket{\tilde{\Psi}_n \vert \Psi(t)}
   \right)
\end{equation}
While the EOMCC excited-state populations are manifestly real, they are
neither bounded above by 1 nor below by 0.
Lack of proper bounds is common in CC theory, but problems are rarely experienced
in practical calculations as long as the CC state vector is a sufficiently good
approximation to the FCI wave function. This, in turn, is related to the distance
in Hilbert space between the reference determinant and the FCI wave function,
as discussed in more detail by \citeauthor{Kristiansen2020}\cite{Kristiansen2020} for 
TDCC theory.

Consistent with this analysis, we compute the ground-state population via the
ground-state projector
\begin{equation}
   \hat{P}_0 =
   \begin{pmatrix}
      \ket{\Psi_0}\bra{\tilde{\Psi}_0} & 0 \\
      0 & \ket{\tilde{\Psi}_0}\bra{\Psi_0}
   \end{pmatrix}
\end{equation}
as
\begin{equation}
\label{eq:gs_pop}
    p_0(t) = \bbrakket{S(t) \vert \hat{P}_0 \vert S(t)}
    =
    \Re\left(
       \braket{\tilde{\Psi}(t) \vert \Psi_0}
       \braket{\tilde{\Psi}_0 \vert \Psi(t)}
   \right)
\end{equation}

\subsection{Projection operators from CC linear response theory}

In lieu of explicitly defined excited states in CC response theory,
we investigate the CC linear response (CCLR) projector
\begin{equation}
\label{eq:cclr_projector}
   \hat{P}_n =
   \begin{pmatrix}
      \ket{\Psi_n}\bra{\tilde{\Psi}_n} & 0 \\
      \sket{\breve{\Psi}_n}\bra{\tilde{\Psi}_n} +
      \ket{\tilde{\Psi}_n}\sbra{\breve{\Psi}_n} &
      \ket{\tilde{\Psi}_n}\bra{\Psi_n}
   \end{pmatrix}
\end{equation}
where we have introduced the notation ($f$ and $g$ are arbitrary functions)
\begin{equation}
    \brasket{g \vert f} \equiv \braket{g \vert f}^*, \qquad
    \sbraket{f \vert g} \equiv \braket{f \vert g}^* =
    \braket{g \vert f} = \brasket{g \vert f}^*
\end{equation}
The functions $\ket{\Psi_n}$ and $\bra{\tilde{\Psi}_n}$
are defined in eq.~\eqref{eq:eom_rl_states}, and
\begin{equation}
\label{eq:breve_psi_n}
   \bra{\breve{\Psi}_n} = \bra{\bar{\Psi}_n} - {\cal R}^n_0 \bra{\tilde{\Psi}_0}
\end{equation}
where
\begin{equation}
   \bra{\bar{\Psi}_n} =
   \bra{\Phi_0} \bar{M}_n \ee^{-T_0} -
   \bra{\tilde{\Psi}_0} R_n, \qquad
   \bar{M}_n = \sum_\mu \bar{\cal{M}}^n_\mu Y_\mu^\dagger
\end{equation}
The amplitudes $\bar{\cal{M}}^n_\mu$ are determined by
the linear equations~\cite{Christiansen1998}
\begin{equation}
\label{eq:Mbar_eqs}
   \left( \boldsymbol{A}^T + \Delta E_n\boldsymbol{1}\right) \bar{\cal M}^n =
    - \boldsymbol{F{\cal R}}^n
\end{equation}
where the superscript $T$ denotes matrix transposition, 
$\boldsymbol{1}$ is the unit matrix, $\boldsymbol{A}$
is the CC Jacobian matrix defined in eq.~\eqref{eq:cc_jacobian},
$\Delta E_n$ is the eigenvalue (excitation energy) 
corresponding to the right eigenvector $\boldsymbol{{\cal R}}^n$
(cf. eq.~\eqref{eq:eigen}), and the symmetric matrix
$\boldsymbol{F}$ is defined by
\begin{equation}
\label{eq:F_matrix}
   F_{\mu\nu} = \braket{\tilde{\Psi}_0 \vert [[H_0,X_\mu],X_\nu] \vert \Psi_0}
\end{equation}

The main arguments in favor of the CCLR projector \eqref{eq:cclr_projector} are
\begin{enumerate}
\item it becomes identical to the EOMCC projector in the FCI limit, and
\item it yields the correct size-intensive CCLR ground-to-excited state transition strength. 
\end{enumerate}
We demonstrate in the appendix that $\bra{\breve{\Psi}_n} = 0$ in the FCI limit,
implying that the CCLR projector \eqref{eq:cclr_projector} becomes identical
to the EOMCC projector \eqref{eq:conventional_eom_projector} in this limit.
The CCLR projector \eqref{eq:cclr_projector} gives the 
correct transition strength
\begin{alignat}{2}
   \bbrakket{S_0\vert \hat{B}\hat{P}_n\hat{C} \vert S_0}
   =
   &\frac{1}{2} \left(
   \braket{\tilde{\Psi}_0 \vert B \vert \Psi_n^x}
   +
   \braket{\bar{\Psi}_n \vert B \vert \Psi_0}
   \right)
   &&\braket{\tilde{\Psi}_n \vert C \vert \Psi_0}
\nonumber \\
   +
   &\frac{1}{2} \left(
   \braket{\tilde{\Psi}_0 \vert C \vert \Psi_n^x}
   +
   \braket{\bar{\Psi}_n \vert C \vert \Psi_0}
   \right)^*
   &&\braket{\tilde{\Psi}_n \vert B \vert \Psi_0}^*
\end{alignat}
To make the equivalence evident, we note that
in the notation used by \citeauthor{Christiansen1998}~\cite{Christiansen1998}
for ``right'' and ``left'' transition moments $T^B_{n0}$ and $T^B_{0n}$
(see eqs. (5.42) and (5.60) of Ref.~\citenum{Christiansen1998}),
\begin{align}
\label{eq:cclr_rmom}
   \braket{\tilde{\Psi}_0 \vert B \vert \Psi_n^x} +
   \braket{\bar{\Psi}_n \vert B \vert \Psi_0}
   &=
   \sum_\mu \braket{\tilde{\Psi}_0\vert [B,X_\mu] \vert\Psi_0} {\cal R}^n_\mu
   +
   \sum_\mu \bar{{\cal M}}^n_\mu 
       \braket{\tilde{\Phi}_\mu\vert\ee^{-T_0}B\ee^{T_0}\vert\Phi_0}
   \equiv T^B_{0n}
\\
\label{eq:cclr_lmom}
   \braket{\tilde{\Psi}_n \vert B \vert \Psi_0}
   &=
   \sum_\mu {\cal L}^n_\mu
       \braket{\tilde{\Phi}_\mu \vert \ee^{-T_0}B \ee^{T_0}\vert \Phi_0}
   \equiv T^B_{n0}
\end{align}
This allows us to recast the transition strength as
\begin{equation}
\label{eq:transition_strength}
   \bbrakket{S_0\vert \hat{B}\hat{P}_n\hat{C} \vert S_0} =
   \frac{1}{2} \left(
   T^B_{0n}T^C_{n0} + \left( T^C_{0n}T^B_{n0} \right)^*
   \right)
\end{equation}
which is identical to the expression obtained as a residue of the CC linear
response function in eq. (5.49) of Ref.~\citenum{Christiansen1998}.

While the CCLR projectors are hermitian with respect
to the indefinite inner product,
$\bbrakket{S_1 \vert \hat{P}_n \vert S_2} =
\bbrakket{S_2 \vert \hat{P}_n \vert S_1}^*$,
they are \emph{not} proper
projection operators as they are neither orthogonal nor idempotent,
\begin{equation}
\label{eq:cclr_orth}
   \hat{P}_m\hat{P}_n = \delta_{mn}\hat{P}_n +
   \begin{pmatrix}
      0 & 0 \\
      \ket{\tilde{\Psi}_m}\bra{\tilde{\Psi}_n}
      \left(
        \braket{\breve{\Psi}_n\vert \Psi_m} +
        \braket{\breve{\Psi}_m\vert \Psi_n}^*
      \right) & 0
   \end{pmatrix}
\end{equation}
The correction term only vanishes in the FCI limit.
They \emph{do}, however, project onto 
the orthogonal complement of the CC ground state, as
\begin{equation}
   \hat{P}_n \kket{S_0} = \frac{1}{\sqrt{2}}
   \begin{pmatrix}
      \ket{\Psi_n}\braket{\tilde{\Psi}_n\vert \Psi_0} \\
      \ket{\tilde{\Psi}_n}
      \left(
        \braket{\breve{\Psi}_n \vert \Psi_0} +
        \braket{\tilde{\Psi}_0 \vert \Psi_n}
      \right)^*
      +
      \sket{\breve{\Psi}_n} \braket{\tilde{\Psi}_n\vert \Psi_0}
   \end{pmatrix}
   =
   \begin{pmatrix}
      0 \\
      0
   \end{pmatrix}
\end{equation}
due to eq.~\eqref{eq:eom_orth_gs} and
\begin{equation}
   \braket{\breve{\Psi}_n\vert \Psi_0}
   =
   \braket{\bar{\Psi}_n\vert \Psi_0} - {\cal R}^n_0
   =
   - \braket{\tilde{\Psi}_0\vert R_n\vert\Psi_0} - {\cal R}^n_0
   = 0
\end{equation}

Excited-state populations extracted by the CCLR projectors according to
eq.~\eqref{eq:excited_pop} become
\begin{equation}
\label{eq:excited_pop_cclr}
   p_n(t) =
   \Re\left[
   \left(
     \braket{\breve{\Psi}_n \vert \Psi(t)} +
     \braket{\tilde{\Psi}(t)\vert \Psi_n}
   \right)
   \braket{\tilde{\Psi}_n \vert \Psi(t)}
   \right]
\end{equation}
While real, the CCLR population is neither bounded below by 0 nor
bounded above by 1.
The ground-state population is given by eq.~\eqref{eq:gs_pop}.

\subsection{Conservation laws}
\label{subsec:conservation}

As noted in Sec.~\ref{subsec:QD_recap}, exact stationary-state populations are conserved in time intervals
where no external forces act on the particle system.
Conservation laws in the framework of TDCC theory have been discussed at various levels of
detail previously.~\cite{Arponen1987,Pedersen1998,Pigg2012,Pedersen2019,Hansen2019,Skeidsvoll2020}

To this end, we recast the TDCC equations of motion in the Hamiltonian form~\cite{Pedersen2019}
\begin{equation}
\label{eq:tdcc_eom_h}
   \dot{\tau}_\mu = -\ii\frac{\partial {\cal H}}{\partial \lambda_\mu}, \qquad
   \dot{\lambda}_\mu = \ii\frac{\partial {\cal H}}{\partial \tau_\mu}, \qquad
   \mu \geq 0
\end{equation}
where the Hamiltonian function ${\cal H} = {\cal H}(\tau,\lambda,t)$ is defined by
\begin{equation}
   {\cal H} = \braket{\tilde{\Psi} \vert H \vert \Psi}
\end{equation}
Introducing the Poisson-like bracket
\begin{equation}
   \left\{ f, g \right\} = \sum_{\mu \geq 0} \left(
   \frac{\partial f}{\partial \tau_\mu} \frac{\partial g}{\partial \lambda_\mu}
   -
   \frac{\partial g}{\partial \tau_\mu} \frac{\partial f}{\partial \lambda_\mu}
   \right)
\end{equation}
for any analytic functions $f=f(\tau,\lambda,t)$ and $g = g(\tau,\lambda,t)$,
the relation
\begin{equation}
\label{eq:poisson_eom}
   \frac{\dd f}{\dd t} = -\ii \left\{ f,{\cal H} \right\} + \frac{\partial f}{\partial t}
\end{equation}
is readily obtained from the Hamiltonian equations \eqref{eq:tdcc_eom_h}.
The equation of motion \eqref{eq:poisson_eom} allows us to identify constants of the
motion within TDCC theory, including truncated TDCC methods.
If the function $f$ does not depend on time explicitly, 
i.e., if it only depends on time through the amplitudes,
it is conserved if its Poisson-like bracket with ${\cal H}$ vanishes, $\left\{ f, {\cal H}\right\} = 0$.
Since $\left\{ {\cal H}, {\cal H}\right\} = 0$, we have
\begin{equation}
   \frac{\dd {\cal H}}{\dd t} = \frac{\partial {\cal H}}{\partial t}
   = \braket{\tilde{\Psi} \vert \frac{\partial H}{\partial t} \vert \Psi}
\end{equation}
which shows that energy is conserved whenever the Hamiltonian operator is constant in time,
including before and after the application of external forces such as laser pulses.
Note that this is true regardless of the truncation of the cluster operators and regardless of
the initial conditions of the amplitudes. This observation was used in Ref.~\citenum{Pedersen2019}
to propose symplectic numerical integration as a stable method for solving the TDCC equations of motion.

The time evolution of the TDCC expection value in eq.~\eqref{eq:expval_cc} is obtained by
choosing $f = \braket{\tilde{\Psi} \vert C \vert \Psi}$. Using a derivation analogous to
that of \citeauthor{Skeidsvoll2020}~\cite{Skeidsvoll2020}, we find
\begin{equation}
   f = \braket{\tilde{\Psi} \vert C \vert \Psi} \quad \implies \quad
   \left\{ f,{\cal H} \right\} = \braket{\tilde{\Psi} \vert [C, H]_\parallel \vert \Psi}
\end{equation}
where we have introduced the projected commutator
\begin{equation}
   [C, H]_\parallel =  C \ee^T P_\parallel \ee^{-T} H -  H \ee^T P_\parallel \ee^{-T} C
\end{equation}
with
\begin{equation}
   P_\parallel = \ket{\Phi_0}\!\bra{\Phi_0} + \sum_\mu \ket{\Phi_\mu}\!\bra{\tilde{\Phi}_\mu}
\end{equation}
Here, the summation over excited determinants is truncated at the same level as the cluster operators.
The time evolution of the TDCC expectation value thus is
\begin{equation}
\label{eq:expval_evolution}
   \frac{\dd}{\dd t} \braket{C} =
   -\ii \frac{1}{2} \left(
   \braket{\tilde{\Psi} \vert [C,H]_\parallel \vert \Psi}
   -
   \braket{\tilde{\Psi} \vert [C^\dagger,H]_\parallel \vert \Psi}^*
   \right)
   +
   \braket{\frac{\partial C}{\partial t}}
\end{equation}
This is not quite the form of a generalized Ehrenfest theorem since,
in general, $[C^\dagger,H]_\parallel \neq -[C,H]_\parallel^\dagger$.
Consequently, constants of the motion in exact quantum dynamics are not
necessarily conserved in truncated TDCC theory.
In the FCI limit, however, $P_\parallel = 1$ and the Ehrenfest theorem
\begin{equation}
\label{eq:ehrenfest}
   \frac{\dd}{\dd t} \braket{C} =
   -\ii\braket{[C, H]} + \braket{\frac{\partial C}{\partial t}}
\end{equation}
is recovered from eq.~\eqref{eq:expval_evolution}.

The EOMCC stationary-state population, eq.~\eqref{eq:excited_pop_eom_conventional},
is of the form \eqref{eq:expval_cc} with $C = \ket{\Psi_n}\!\bra{\tilde{\Psi}_n}$ and
the time evolution, therefore,
is given by eq.~\eqref{eq:expval_evolution}.
The proposed EOMCC projector thus breaks the conservation law of stationary-state populations
in truncated TDCC simulations, although we note that it is properly restored in the
FCI limit, since $C = \ket{\Psi_n}\!\bra{\tilde{\Psi}_n}$
commutes with $H_0$ according to eqs.~\eqref{eq:tise_cc_r} and \eqref{eq:tise_cc_l}
in the Appendix. It seems reasonable to expect that the conservation law is approximately
fulfilled whenever the many-electron dynamics predominantly involves stationary states that
are well approximated within (truncated) EOMCC theory.

The CCLR stationary-state population, eq.~\eqref{eq:excited_pop_cclr}, is not of the form
\eqref{eq:expval_cc}. Instead, the time evolution is given by eq.~\eqref{eq:poisson_eom}
with
\begin{equation}
   f =
   \left(
     \braket{\breve{\Psi}_n \vert \Psi(t)} +
     \braket{\tilde{\Psi}(t)\vert \Psi_n}
   \right)
   \braket{\tilde{\Psi}_n \vert \Psi(t)}
\end{equation}
which only depends on time through the amplitudes ($\partial f/\partial t = 0$).
Hence, also the CCLR projector breaks the conservation law of stationary-state populations
in truncated TDCC simulations. It is restored in the FCI limit where the CCLR and EOMCC
projectors are identical. Again, it seems reasonable to expect that the conservation law
is approximately fulfilled whenever CCLR theory provides a sufficiently good approximation to the FCI
states.

\section{Computational details}
\label{sec:computational_details}

Explicitly time-dependent simulations are performed with a 
closed-shell spin-restricted TDCCSD Python code 
generated using a locally modified version of
the Drudge/Gristmill suite for symbolic 
tensor algebra developed by Zhao and Scuseria~\cite{Drudge_Gristmill}.
The static HF reference orbitals and Hamiltonian integrals are computed
by the Dalton quantum chemistry program~\cite{Aidas2014,Olsen2020}
along with the response vectors required for the EOMCC and CCLR projectors
using the implementations described in
Refs.~\citenum{Koch1996,Christiansen1996,Halkier1997,Christiansen1998a}.
Tight convergence criteria are employed: $10^{-10}\,\au$
for the HF orbital gradient norm (implying machine precision for the
HF ground-state energy), and $10^{-8}\,\au$ for the
CCSD residual norms (both ground state and response equations).
Dunning's correlation-consistent basis
sets~\cite{Dunning1989,Kendall1992,Woon1994},
downloaded from the Basis Set Exchange~\cite{Pritchard2019}, are used
throughout. We also perform TDFCI simulations using the contraction
routines implemented in the PySCF package~\cite{Sun2018}.
The TDCCSD and TDFCI equations of motion are integrated using the
symplectic
Gauss-Legendre integrator~\cite{HairerLubichWanner_GNI} as described
in Ref.~\citenum{Pedersen2019}. 

To test the proposed EOMCC and CCLR projectors, TDCCSD and
TDFCI simulations are carried out for the He and Be atoms placed at the
coordinate origin.
Further tests are performed for the LiH molecule placed on the $z$-axis with
the Li atom at the origin and the H atom at $z=3.08\,\au$,
and for the CH$^+$ ion placed on the $z$-axis with the C atom at the origin and the H atom
at $z=2.13713\,\au$ Finally, we study the time evolution of stationary-state populations
during the optical pump pulse applied by \citeauthor{Skeidsvoll2020}~\cite{Skeidsvoll2020}
to investigate transient x-ray spectroscopy of LiF. As in Ref.~\citenum{Skeidsvoll2020}, we
place the LiF molecule
on the $z$-axis with the F atom at the origin and the Li atom at
$z=-2.9552749018\,\au$
All electrons are correlated and point group symmetry is not
exploited in these simulations.

We assume the systems are intially in the ground state and expose them
to a laser pulse described by the semi-classical interaction operator
in the electric dipole approximation,
\begin{equation}
\label{eq:Vt}
    V(t) = -\boldsymbol{d}\cdot\boldsymbol{u}{\cal E}(t)
\end{equation}
where $\boldsymbol{d}$ is the electric dipole operator of the electrons,
$\boldsymbol{u}$ is the real unit polarization vector of the electric field,
and ${\cal E}(t)$ is the time-dependent electric-field amplitude.

Two forms of the electric-field amplitude are used in this work.
One is the sinusoidal pulse
\begin{equation}
   {\cal E}(t) =
      {\cal E}_0 
      \sin(\omega_0 (t-t_0) + \phi(t)) G(t)
\end{equation}
where ${\cal E}_0$ is the field strength,
$\omega_0$ is the carrier frequency of the pulse,
and $t_0$ is the time at which the pulse is turned on.
The time-dependent phase $\phi(t)$ may effectively
alter the instantaneous carrier frequency, creating a chirped 
laser pulse, if it depends on time at least
quadratically~\cite{Diels_Rudolph_UltrashortLaserPulsePhenomena}.
In this work, we use the quadratic form,
\begin{equation}
   \phi(t) = a(t-t_0) + b(t-t_0)^2
\end{equation}
which, for $b \neq 0$, creates a linearly chirped laser pulse with
instantaneous frequency $\omega(t) = \omega_0 + a + b(t-t_0)$.
The envelope $G(t)$ controls the shape and duration of the pulse
and is defined by
\begin{equation}
\label{eq:sin2_envelope}
   G(t) = \sin^2\left( \pi \frac{t-t_0}{t_d} \right)
          \Theta(t-t_0) \Theta(t_d-(t-t_0))
\end{equation}
where $t_d$ is the duration of the pulse and $\Theta(t)$ is the Heaviside
step function. 

The second form of the electric-field amplitude used in this work
is the Gaussian pulse
\begin{equation}
   {\cal E}(t) =
      {\cal E}_0 
      \cos(\omega_0 (t-t_0)) G(t)
\end{equation}
with the envelope
\begin{equation}
   G(t) = \ee^{-(t-t_0)^2/2\sigma^2} \Theta(t-(t_0-N\sigma))\Theta((t_0+N\sigma)-t)
\end{equation}
where $t_0$ is the central time of the pulse and $\sigma$ is the Gaussian root-mean-square (RMS) width,
and $N$ defines the start time ($t_0-N\sigma$) and end time ($t_0+N\sigma$)
of the pulse through the Heaviside step functions. Note that $N$ thus introduces discontinuities
of the Gaussian pulse at each end. Unless a sudden disturbance of the system is intended, one must
choose $N$ large enough that the discontinuities are negligible.

\section{Results and Discussion}
\label{sec:results_and_discussion}

\subsection{Excited-state Rabi oscillations}
\label{subsec:excited_state_rabi_oscillations}

Involving population inversion,
Rabi oscillations between the ground state and an excited state are very hard to
simulate within conventional TDCC theory with a static HF reference
determinant.
At the periodically recurring points in time where the ground-state population
vanishes, the weight of the HF determinant becomes very small or zero,
making the numerical integration of the TDCC equations exceedingly
challenging.~\cite{Pedersen2019,Kristiansen2020}
Dynamic orbitals, such as those of orbital-adaptive TDCC (OATDCC) theory~\cite{Kvaal2012},
are required for a numerically stable integration of the equations of
motion.~\cite{Kristiansen2020} Consequently, in this work, we will focus on Rabi
oscillations between excited states.

Rabi oscillations between two excited states can be achieved by application of
two consecutive laser pulses, the first of which is resonant with a dipole-allowed
transition from the ground state while the second is resonant with a dipole-allowed
transition between the resulting excited state and another one. The intensity and duration of the
first pulse must be such that the ground-state population is significantly reduced but
not entirely depleted. Nonlinear optical processes
are thus involved, making it an
ideal test case for the CCLR and EOMCC projectors, which are constructed on the basis
of first-order perturbative arguments (first-order perturbation theory in the case of CCLR and
linearization of the cluster exponential in the case of EOMCC), which can not
necessarily be expected to correctly capture higher-order optical processes.
In particular, transition moments between excited states are quadratic response
properties, which can not be expressed solely in terms of linear response
parameters~\cite{Christiansen1998}. It is, therefore, important to test
if the proposed projectors correctly capture the effects of nonlinear
optical processes.

\begin{figure}
  \includegraphics{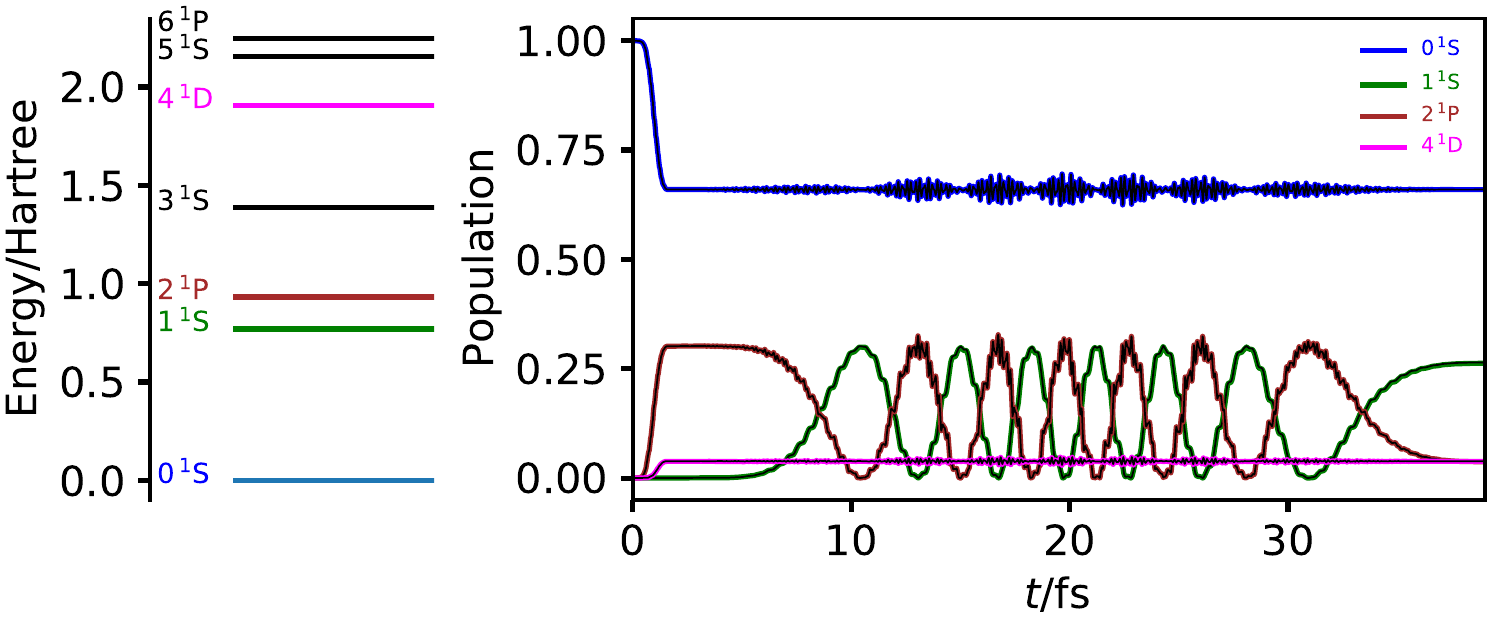}
  \caption{Energy-level populations computed with the aug-cc-pVTZ basis set through CCLR projectors 
           plotted as functions of time for He exposed to two consecutive laser pulses, the first
           resonant with the $0^1\text{S} \to 2^1\text{P}$ transition and the second resonant
           with the $2^1\text{P} \to 1^1\text{S}$ transition, with peak intensity 
           $87.7\,\text{TW/cm}^2$.
           The black curves show the populations during the TDFCI simulation.
          }
  \label{fig:he_avtz_pop}
\end{figure}

Results for the He atom with the aug-cc-pVTZ basis set
are presented in Fig.~\ref{fig:he_avtz_pop}.
The integration of the TDCCSD equations of motion was
performed with time step $\Delta t = 0.1\,\au = 2.42\,\text{as}$
using the eighth-order ($s=4$) Gauss-Legendre integrator and a convergence
threshold of $10^{-10}$ (residual norm) for the fixed-point iterations.
The ground- and excited-state energy levels shown in the left panel of Fig.~\ref{fig:he_avtz_pop}
were computed using CCSD linear response theory.
In total, $14$ excited states were computed, several of which lie above the ionization
energy, which is estimated to be $0.902\,\au$ using the total ground-state energy difference
between the neutral and ionized atom at the (spin unrestricted)
CCSD/aug-cc-pVTZ level of theory~\cite{CCCBDB}.
Although the states above the ionization energy are unphysical, we keep them for the purpose
of comparing with regular TDFCI simulations with the same basis set.
The first sinusoidal laser pulse, applied at $t_0 = 0\,\au$,
is resonant with the $0^1\text{S} \to 2^1\text{P}$ transition
at $\omega_0 = 0.932\,\au$ and has peak intensity $87.7\,\text{TW/cm}^2$
(${\cal E}_0 = 0.05\,\au$) with a duration of $10$ optical cycles
($t_d = 67.41\,\au = 1.63\,\text{fs}$). The second sinusoidal pulse, applied immediately
after the first pulse, is resonant with the $2^1\text{P} \to 1^1\text{S}$ transition
at $\omega_0 = 0.163\,\au$ at the same peak intensity as the first pulse
with a duration of $40$ optical cycles ($t_d = 1545.19\,\au = 37.38\,\text{fs}$).
Both lasers are polarized along the $z$-axis with constant phase $\phi(t) = 0$.
The ground and excited levels are dominated by the electron configurations
$1\text{s}^2$ (ground state, $0^1\text{S}$), $1\text{s}^12\text{s}^1$ ($1^1\text{S}$),
and $1\text{s}^12\text{p}^1$ ($2^1\text{P}$).

The right panel of Fig.~\ref{fig:he_avtz_pop} shows the total energy-level populations
computed using the CCLR projector
as a function of time during the application of the two consecutive laser pulses.
Energy levels that are never populated above $0.01$ are excluded.
The level populations are computed by summing up populations of all states belonging to each 
energy level, thus avoiding ambiguities arising from the arbitrariness of the basis of a
degenerate subspace.
As expected, the
first laser pulse causes significant population of the $2^1\text{P}$ energy level.
The high-lying $4^1\text{D}$ level, which is dominated by the $1\text{s}^13\text{d}^1$ electron
configuration and located $0.972\,\au$ above the $2^1\text{P}$ level,
also becomes populated towards the end of the first pulse
due to the dipole-allowed transition $2^1\text{P} \to 4^1\text{D}$.
The length-gauge 
oscillator strength of this transition is $f=0.484$ compared with $f=0.355$ for the
$0^1\text{S} \to 2^1\text{P}$ transition. The population of the $4^1\text{D}$ level is still
modest, however, since the transition can only occur once the $2^1\text{P}$ level is 
significantly populated.

The second pulse induces several cycles of Rabi oscillations between the $2^1\text{P}$
and $1^1\text{S}$ levels. The Rabi oscillations are slightly perturbed by weak
transitions between the $2^1\text{P}$ level and the $0^1\text{S}$ ground state, as witnessed by
the increasing oscillation of the ground-state population when the population of the
$2^1\text{P}$ level is close to its maximum value. We also observe an even weaker
perturbation caused by the higher-lying $4^1\text{D}$ level.
The CCLR populations agree both with TDFCI populations
and with EOMCC populations: the RMS deviation
for the entire simulation is $10^{-3}$ between the CCLR and TDFCI populations
and $3 \times 10^{-7}$ between the CCLR and EOMCC populations. 
We have previously demonstrated that discrepancies
with TDFCI simulations for the He atom can be reduced by tightening computational
parameters such as convergence thresholds and, most importantly, by reducing
the time step of the numerical integration~\cite{Kristiansen2020}.
We thus ascribe the small discrepancies between the CCLR and TDFCI populations
to the rather coarse discretization ($\Delta t = 0.1\,\au$)
employed in the numerical integration scheme, and
conclude that the
proposed CCLR and EOMCC
projectors behave correctly in the FCI limit.

It is of interest to compare the TDCCSD simulation with a much simpler model
based on an eigenstate expansion propagated according to eq.~\eqref{eq:tdse_model}.
Letting $\ket{n}, n=0,1,2,\cdots,14$, represent the stationary states computed
with CCSD linear response theory, all we need to integrate eq.~\eqref{eq:tdse_model}
in the presence of external laser pulses of the form \eqref{eq:Vt}
is the dipole matrix in the energy eigenbasis.
This model is essentially identical to that employed by Sonk et al.~\cite{Sonk2011},
except that we use CCSD linear and quadratic response theory
rather than EOMCCSD theory to build the dipole matrix.
(Note, however, that the CCSD response and EOMCCSD approaches yield identical
results for He.) It is also similar to the EOMCCSD model employed by
Luppi and Head-Gordon~\cite{Luppi2012}, who propagated both bra and ket states.

The only obstacle is that the ``right'' and ``left'' transition moments from
CC response theory, cf.
eqs.~\eqref{eq:cclr_rmom} and \eqref{eq:cclr_lmom},
are not related by complex conjugation, yielding a spurious nonhermitian
dipole matrix. Sonk et al.~\cite{Sonk2011} and Luppi and Head-Gordon~\cite{Luppi2012}
circumvented this issue by only using the hermitian part of the matrix.
In the present
case, due to symmetry, the CCSD electric-dipole transition moments are either zero
or may be chosen parallel to one of the three Cartesian axes, making it easy to define
the off-diagonal dipole matrix
elements as the negative square root of the dipole transition strength (cf.
eq.~\eqref{eq:transition_strength} with $B=C$ a Cartesian component of the
position operator). With the dipole matrix thus constructed for the $14$ states, we have integrated
eq.~\eqref{eq:tdse_model}
with the initial condition $C_n(t=0) = \delta_{n0}$, and the same consecutive
laser pulses as in the He/aug-cc-pVTZ TDCCSD simulations. The Gauss-Legendre
integrator was used with the same parameters and time step.

The RMS population deviation between the model and the full TDCCSD simulations
is about twice that between the TDFCI and TDCCSD simulations.
The maximum absolute deviation is an order of magnitude greater ($0.02$ versus $0.002$), however, indicating that
a potentially large number of states, including states above the ionization energy,
may be required for the simple model to agree quantitatively with full TDCCSD simulations in general.

\begin{figure}
  \includegraphics{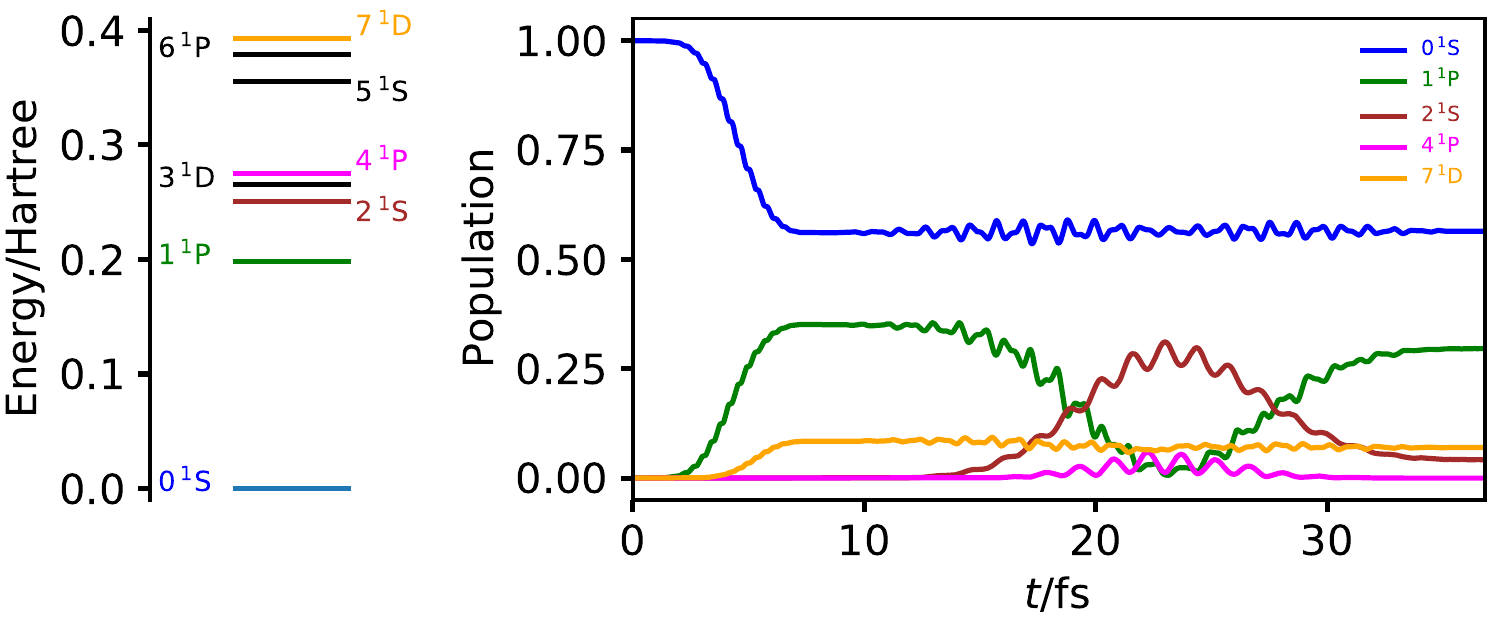}
  \caption{Energy-level populations computed with the aug-cc-pVDZ basis set through CCLR projectors 
           plotted as functions of time for Be exposed to two consecutive laser pulses, the first
           resonant with the $0^1\text{S} \to 1^1\text{P}$ transition and the second resonant
           with the $1^1\text{P} \to 2^1\text{S}$ transition, with peak intensity 
           $0.877\,\text{TW/cm}^2$.}
  \label{fig:be_avdz_pop}
\end{figure}

Moving away from the FCI limit, we repeat the study of excited-state Rabi oscillations for
the Be atom with the aug-cc-pVDZ basis set. The integration parameters were the same as
those used for He above.
The ground- and excited-state energy levels shown in the left panel of Fig.~\ref{fig:be_avdz_pop}
were computed using CCSD linear response theory. 
In total, $21$ excited states were computed and the CCSD excitation energies agree with
those of FCI theory to within $3\times 10^{-4}\,\au$.
Several of the computed excited states lie above the ionization
energy, which is estimated to be $0.341\,\au$ using the total ground-state energy difference
between the neutral and ionized atom at the (spin unrestricted)
CCSD/aug-cc-pVDZ level of theory~\cite{CCCBDB}.
The high-lying excited states are retained in order to compare
with regular TDFCI simulations with the same basis set.
The first sinusoidal laser pulse, applied at $t_0 = 0\,\au$,
is resonant with the $0^1\text{S} \to 1^1\text{P}$ transition
at $\omega_0 = 0.198\,\au$ and has peak intensity $0.877\,\text{TW/cm}^2$
(${\cal E}_0 = 0.005\,\au$) with a duration of $10$ optical cycles 
($t_d = 316.93\,\au = 7.64\,\text{fs}$). The second sinusoidal pulse, applied immediately
after the first pulse, is resonant with the $1^1\text{P} \to 2^1\text{S}$ transition
at $\omega_0 = 0.0522\,\au$ at the same peak intensity as the first pulse
with a duration of $10$ optical cycles ($t_d = 1204.59\,\au = 29.14\,\text{fs}$).
Both lasers are polarized along the $z$-axis with constant phase $\phi(t) = 0$.
The ground and excited levels are dominated by the electron configurations
$1\text{s}^22\text{s}^2$ (ground state, $0^1\text{S}$), $1\text{s}^22\text{s}^12\text{p}^1$ ($1^1\text{P}$),
and $1\text{s}^12\text{s}^13\text{s}^1$ ($2^1\text{S}$).

The right panel of Fig.~\ref{fig:be_avdz_pop} shows the total energy-level populations
computed using the CCLR projector
as a function of time during the application of the two consecutive laser pulses.
Energy levels that are never populated above $0.01$ are excluded.
The
first laser pulse causes significant population of the $1^1\text{P}$ energy level
by excitation from the ground state (oscillator strength $f=0.478$),
although the transition is quenched by
further excitation to the
high-lying $7^1\text{D}$ level from the $1^1\text{P}$ level
($f = 0.685$). The $7^1\text{D}$ level,
which is dominated by the $1\text{s}^22\text{s}^13\text{d}^1$ 
electron configuration, is located $0.195\,\au$ above the $1^1\text{P}$ level and,
hence, the $1^1\text{P} \to 7^1\text{D}$ transition is nearly resonant with the first
laser. Consequently, the $7^1\text{D}$ level population increases
once sufficient population of the $1^1\text{P}$ level is achieved towards
the end of the first pulse.
The second pulse induces a single-cycle Rabi oscillation between the $1^1\text{P}$
and $2^1\text{S}$ levels ($f=0.118$), which is quite significantly perturbed by 
the transition between the $2^1\text{S}$ and $4^1\text{P}$ levels ($f=0.211$).
The population of the $4^1\text{P}$ level drops to zero as the Rabi oscillation
enters the final stage where the population of the $2^1\text{S}$ level decreases.

The CCLR populations are in close agreement with TDFCI populations
and with EOMCC populations: the RMS deviation
for the entire simulation is $7.4 \times 10^{-3}$ between the CCLR and TDFCI populations
and $9.1 \times 10^{-5}$ between the CCLR and EOMCC populations.

\begin{figure}
  \includegraphics{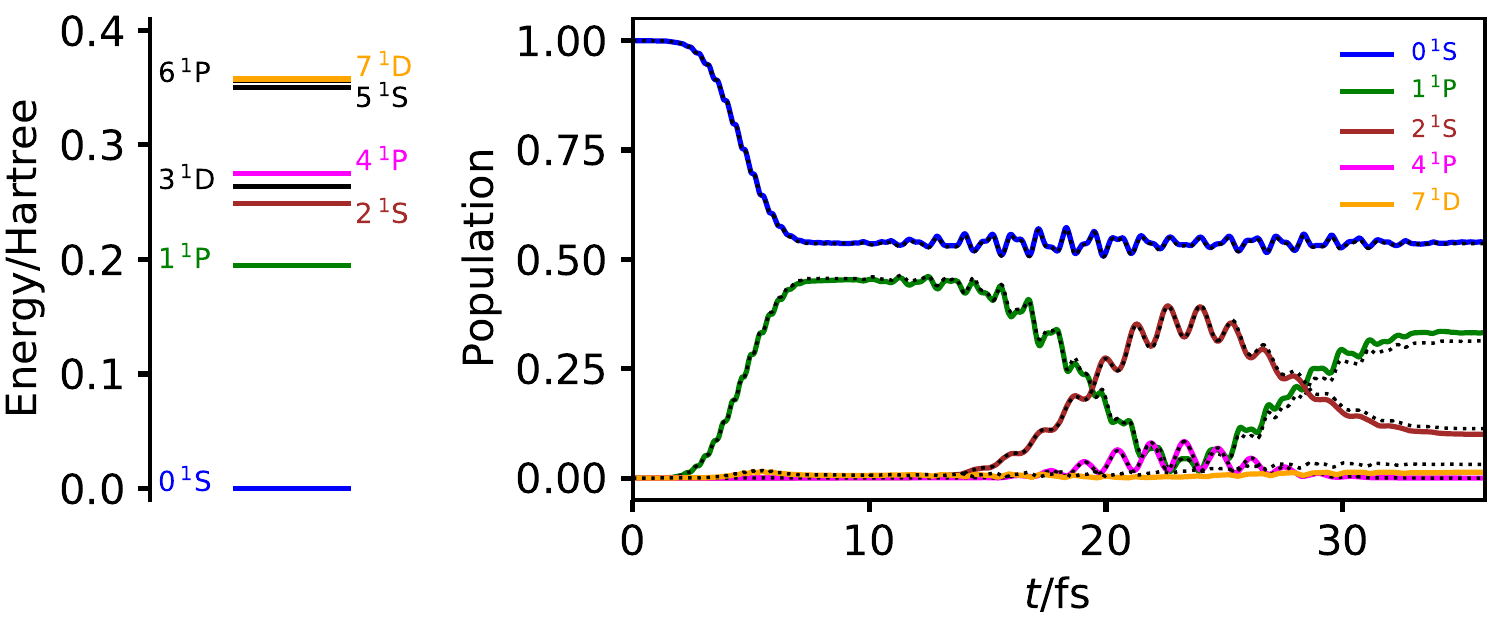}
  \caption{Energy-level populations computed with the aug-cc-pVTZ basis set through CCLR projectors 
           plotted as functions of time for Be exposed to two consecutive laser pulses, the first
           resonant with the $0^1\text{S} \to 1^1\text{P}$ transition and the second resonant
           with the $1^1\text{P} \to 2^1\text{S}$ transition, with peak intensity 
           $0.877\,\text{TW/cm}^2$. The dotted lines show the populations obtained using a
           simplified eigenstate expansion (see text) with the $22$ states forming the $8$ 
           energy levels of the left panel included.}
  \label{fig:be_avtz_pop}
\end{figure}

Increasing the basis set to aug-cc-pVTZ, the CCSD levels, the higher-lying ones in particular,
move down in energy as 
seen by comparing the left panel of Fig.~\ref{fig:be_avtz_pop} with that of
Fig.~\ref{fig:be_avdz_pop}.
The right panel of Fig.~\ref{fig:be_avtz_pop} shows the variation of the level populations
as the Be atom is exposed to the same sinusoidal laser pulses as in Fig.~\ref{fig:be_avdz_pop},
albeit with the carrier frequencies adjusted to match the $0^1\text{S} \to 1^1\text{P}$
and $1^1\text{P} \to 2^1\text{S}$ transitions at $\omega_0 = 0.195\,\au$ and
$\omega_0 = 0.0539\,\au$, respectively. The duration is $10$ optical cycles
for each pulse, as above.
The populations obtained from the CCLR and EOMCC projectors
are vitually identical, with an overall RMS deviation of $1.6 \times 10^{-4}$.

The lowering of the $7^1\text{D}$ level, which is now $0.162\,\au$ above the
$1^1\text{P}$ level, implies that the probability of the $1^1\text{P} \to
7^1\text{D}$ transition ($f=0.648$) diminishes, resulting in very low population of the
$7^1\text{D}$ level and increased population of the $1^1\text{P}$ level
(compared with the aug-cc-pVDZ simulation) during the first pulse.
Although the populations of the states involved thus are different,
the perturbed Rabi oscillation induced by the second pulse
is essentially the same as in Fig.~\ref{fig:be_avdz_pop}.

While a full TDFCI simulation is too costly with the aug-cc-pVTZ basis set,
we compare with the much simpler model
introduced for He above.
We use the same consecutive
laser pulses and the Gauss-Legendre integrator with the same parameters
as in the Be/aug-cc-pVTZ TDCCSD simulation 
for the model simulation with $22$ states included.

The resulting energy-level
populations, plotted as dotted lines in Fig.~\ref{fig:be_avtz_pop},
are remarkably similar to the full TDCCSD results.
The maximum absolute deviations between the model and TDCCSD populations
are just $15\%$ for the $1^1\text{P}$ level, $2\%$ for the $2^1\text{S}$ and
$7^1\text{D}$ levels, and below $1\%$ for the remaining levels, including the
ground state. Such good results can only be expected from the simple model when
all, or very nearly all, participating CCSD states are included. 
How many states are needed will in general be very hard to estimate \emph{a priori}.

\subsection{Control by chirped laser pulses}

\begin{figure}
  \includegraphics{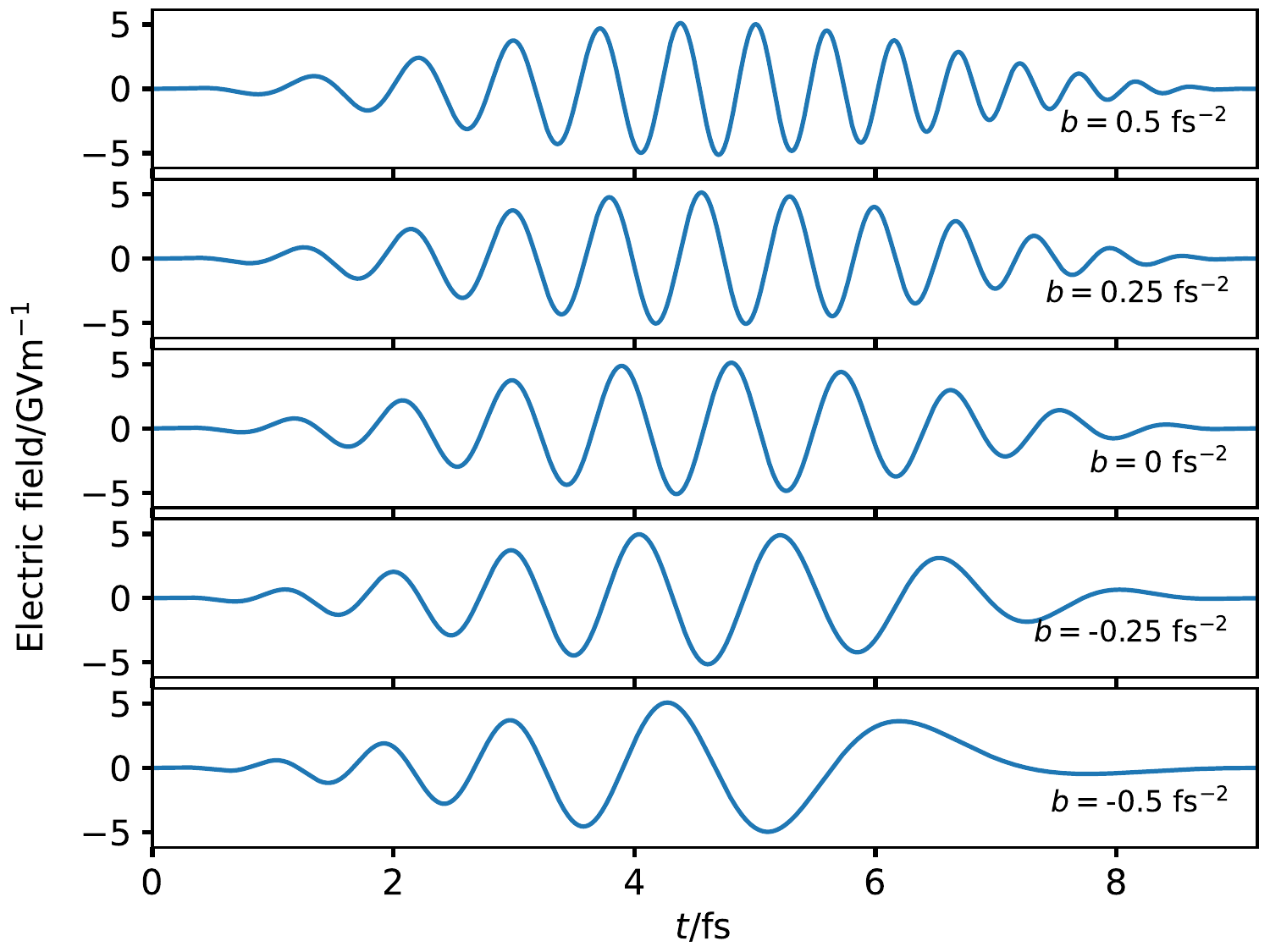}
  \caption{Laser pulses with different chirp rates.}
  \label{fig:chirped_field}
\end{figure}

Control by shaped laser pulses is an important challenge to theoretical simulations
and requires information about population of energy levels.
For the LiH molecule described with the aug-cc-pVDZ basis set, which is small
enough to allow TDFCI reference simulations, we use chirped sinusoidal laser pulses to further test the
proposed CCLR and EOMCC projectors within TDCCSD simulations. The laser pulses are polarized
along the $x$-axis, perpendicular to the molecular axis,
and the duration is kept fixed at $t_d=378.4\,\au = 9.152\,\text{fs}$, corresponding to $10$ optical
cycles of radiation resonant with the lowest-lying electric-dipole allowed transition
from the ground state, the $x$-polarized $^1\Sigma^+ \to \,^1\Pi$ transition
at $\omega_0=0.166\,\au$. The laser pulses are turned on at $t_0 = 0\,\au$ with peak
intensity $3.51\,\text{TW/cm}^2$ (${\cal E}_0 = 0.01\,\au$).
The oscillator strength of this transition is estimated to be $f=0.208$ by CCSD linear
response theory with the aug-cc-pVDZ basis set. The phase of the laser pulse is defined
such that the instantaneous frequency is
\begin{equation}
    \omega(t) = \omega_0 + b\left( t - \frac{t_d}{3}\right)
\end{equation}
and the chirp rate $b$ is varied between $-0.513\,\text{fs}^{-2}$ and $+0.513\,\text{fs}^{-2}$.
A few such laser pulses with different chirp rates are shown in Fig.~\ref{fig:chirped_field}.

\begin{figure}
  \includegraphics{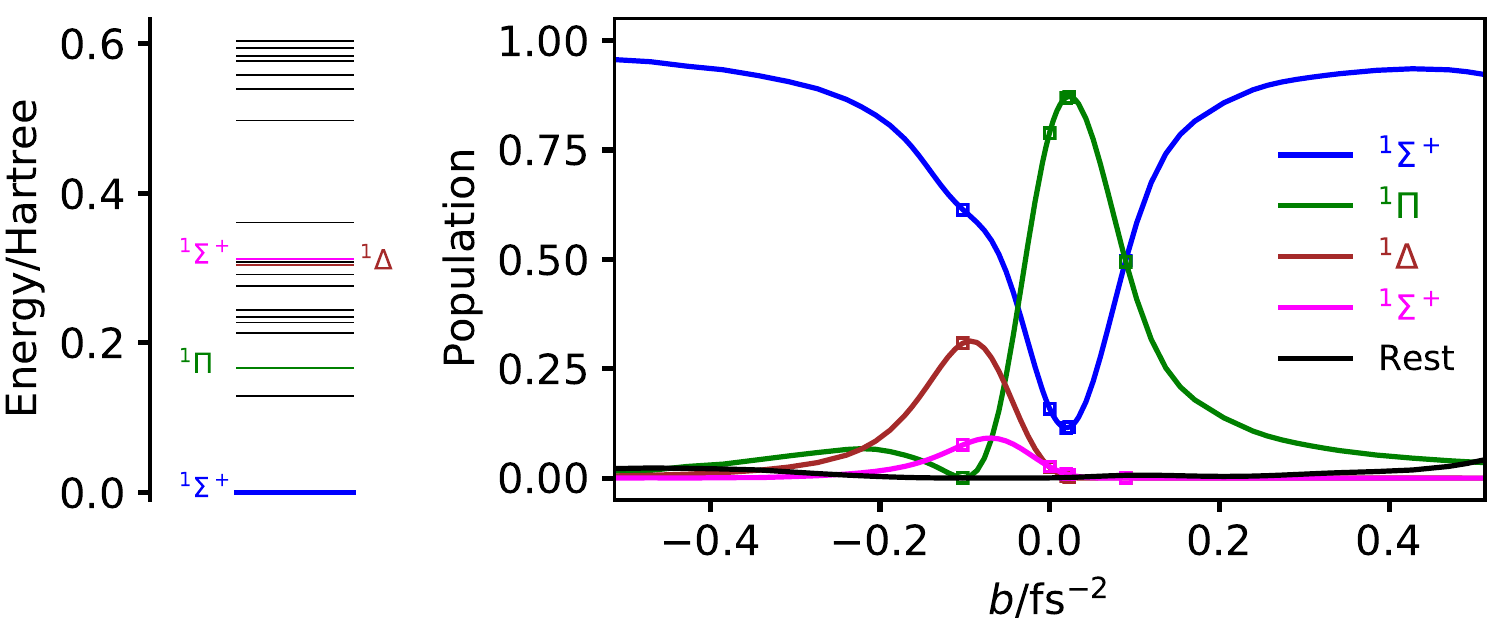}
  \caption{Final population of CCSD energy levels, computed with the CCLR projector,
as a function of chirp rate for LiH with the aug-cc-pVDZ basis set. The squares mark reference populations from TDFCI simulations.}
  \label{fig:lih_avdz_final_pop}
\end{figure}

The $31$ lowest-lying states,
organized into $21$ energy levels in the left panel of Fig.~\ref{fig:lih_avdz_final_pop},
were computed with CCSD linear response theory
(aug-cc-pVDZ basis) and used to construct CCLR and EOMCC projectors for the simulations.
The highest-lying energy level is $0.603\,\au$ above the ground state, well beyond the ionization
energy estimated by CCSD/aug-cc-pVDZ total-energy difference to be $0.281\,\au$~\cite{CCCBDB}.
The most important energy levels in the simulations are marked by their term symbols. These states
are all predominantly single-excited states, with at least $90\%$ contribution from singles in the
EOMCC excitation amplitudes. While the $^1\Pi$ level at $0.166\,\au$ is well below the 
estimated ionization energy, the $^1\Delta$ and $^1\Sigma^+$ levels at
$0.291\,\au$ and $0.312\,\au$, respectively, are slightly above. With $x$-polarized laser pulses,
one-photon transitions from the ground state to the $^1\Delta$ and $^1\Sigma^+$ levels are 
electric-dipole forbidden, implying that these excited levels can only become populated
by nonlinear optical processes.

The final populations of these levels, computed immediately after
the interaction with the chirped sinusoidal laser pulse,
are shown in the right panel of Fig.~\ref{fig:lih_avdz_final_pop} along with a few reference
TDFCI results.
The TDCCSD (and TDFCI) equations of motion were integrated using the sixth order ($s=3$)
Gauss-Legendre integrator with time step $\Delta t = 0.1\,\au$ and convergence threshold
$10^{-6}$.
The sum of the populations of
the remaining $27$ energy levels is labelled ``Rest'' and is seen to be insignificant for
all but the most up- or down-chirped pulses.
At $b=0\,\text{fs}^{-2}$, the pulse is resonant with the ground-state $^1\Sigma^+ \to
\,^1\Pi$ transition and, therefore, other levels are hardly populated. The maximum
population of the $^1\Pi$ level is observed at a slightly up-chirped pulse
(at $b=0.023\,\text{fs}^{-2}$), which
prevents further excitation from the $^1\Pi$ level to the higher-lying $^1\Sigma^+$ and
$^1\Delta$ levels. As the chirp rate increases, the laser pulse becomes increasingly
off-resonant and the ground-state population increases.

\begin{figure}
  \includegraphics{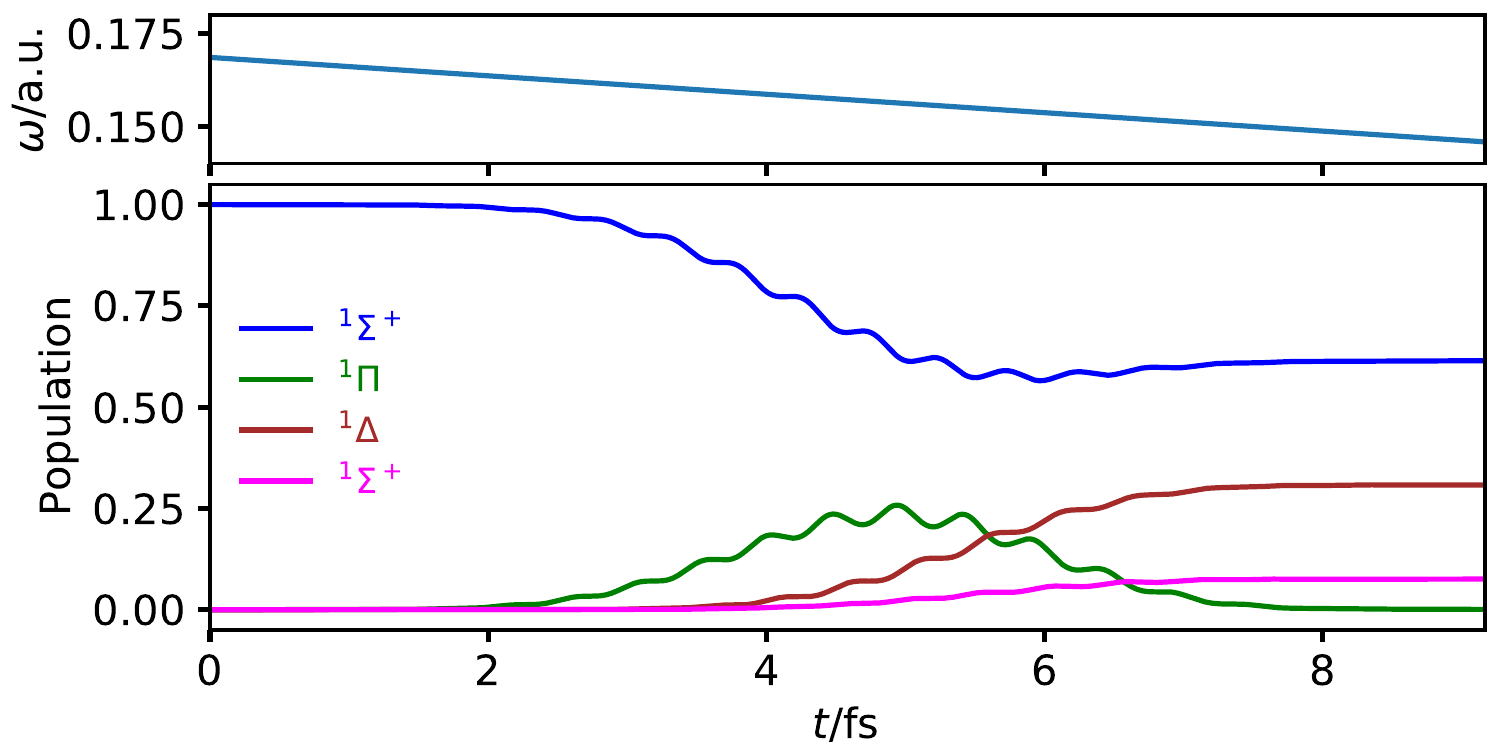}
  \caption{Energy-level populations computed with the aug-cc-pVDZ basis set through CCLR projectors 
           plotted as functions of time for LiH exposed to a down-chirped laser pulse with
           chirp rate $b=-0.102\,\text{fs}^{-2}$. The instantaneous frequency of the laser pulse
           is shown in the top panel.}
  \label{fig:lih_avdz_sim}
\end{figure}

The population of the excited $^1\Sigma^+$ level and, in particular, of the $^1\Delta$ level
increases with moderately down-chirped pulses due to transitions from the $^1\Pi$ level,
whose probability increases as the laser frequency decreases.
At $b=-0.102\,\text{fs}^{-2}$, the population of the $^1\Pi$ level is just $9.3\times 10^{-4}$,
while the excited $^1\Sigma^+$ and $^1\Delta$ populations are close to their maximum
values. These nonlinear optical processes can easily be understood by studying the
populations during interaction with the laser pulse in Fig.~\ref{fig:lih_avdz_sim}.
During the first half of the pulse, the instantaneous laser frequency is nearly resonant
with the ground-state $^1\Sigma^+ \to \,^1\Pi$ transition at $0.166\,\au$ ($f=0.207$),
causing the population of the $^1\Pi$ level to increase.
As the instantaneous frequency decreases, it comes closer to the
transition frequencies of the $^1\Pi \to \,^1\Delta$ and $^1\Pi \to \,^1\Sigma^+$ 
transitions at $0.138\,\au$ and $0.146\,\au$, respectively. Although the instantaneous
frequency is closer to the $^1\Pi \to \,^1\Sigma^+$ transition than to the
$^1\Pi \to \,^1\Delta$ transition, the latter level is considerably more populated.
This is explained by the difference in oscillator strength for these transitions:
$f=0.363$ for the $^1\Pi \to \,^1\Delta$ transition compared with $f=0.076$ for the
$^1\Pi \to \,^1\Sigma^+$ transition.

The differences between the EOMCC and CCLR projectors are, again, utterly
insignificant, the typical RMS population deviation beetween them being approximately $10^{-5}$
regardless of chirp rate.
As can be seen from Fig. \ref{fig:lih_avdz_final_pop}, the TDCCSD simulations are in
excellent agreement with TDFCI results. This is not unexpected, since all states participating
in the dynamics are single-excited states. The maximum absolute deviation in the $30$ 
excitation energies, including the excited states with significant double-excited
character (down to $11\%$ singles contribution in the EOMCC excitation amplitudes),
between CCSD linear response theory and FCI theory is just $0.0013\,\au$.

\subsection{Dynamics involving double-excited states}

In general, CCSD linear response theory performs poorly for states dominated by
double-excited determinants relative to the HF ground state. For such states,
excitation-energy errors are typically an order of magnitude greater than for
single-excited states~\cite{Sauer2009,Bartlett2012}, roughly $0.01\,\au$ for double-excited states 
compared with $0.001\,\au$ for single-excited states
of small molecules where FCI results are available~\cite{Bartlett2012}.
In all examples presented above, the states participating
significantly in the dynamics are all single-excitation dominated, explaining the close
agreement observed between TDFCI and TDCCSD simulations.

With a ground-state wave function dominated by the HF ground-state determinant,
one-photon transitions to excited states dominated by double-excited determinants are either
electric-dipole forbidden or only weakly allowed. Accordingly, we expect double-excited states
to influence laser-driven many-electron dynamics mainly through nonlinear optical processes.
In order to test the influence on TDCCSD dynamics,
we consider the CH$^+$ molecule, which is a classic example of the relatively
poor performance of CCSD linear response and EOMCCSD theory for such states,
see, for example, Refs.~\citenum{Koch1990a,Watts1994,Watts1995}. 

The $^1\Sigma^+$ ground state of CH$^+$ is dominated by the $1\sigma^22\sigma^23\sigma^2$ electron
configuration with some non-dynamical correlation contribution from the double-excited
$1\sigma^22\sigma^21\pi^2$ configuration. The two lowest-lying excited states
form the $^1\Pi$ energy level and are dominated by the single-excited
$1\sigma^22\sigma^23\sigma^11\pi^1$ configuration.
The three subsequent states form two energy levels, $^1\Delta$ and $^1\Sigma^+$, and
are almost purely double-excited states stemming from the $1\sigma^22\sigma^21\pi^2$
electron configuration. Transitions from the ground state to these levels are either
electric-dipole forbidden ($^1\Delta$) or very weak with oscillator strengths on the 
order of $10^{-3}$. Significant population of these levels, therefore, can only
be achieved through nonlinear optical processes, requiring rather intense laser pulses.

\begin{figure}
  \includegraphics{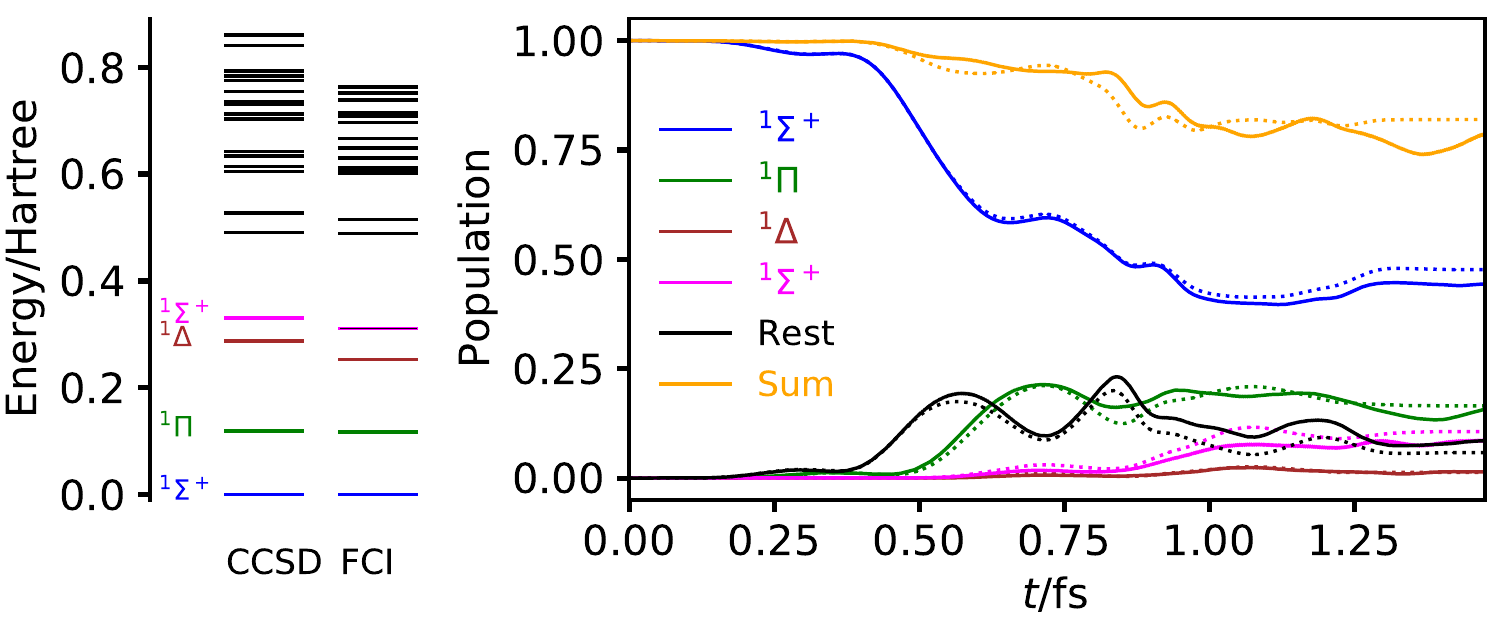}
  \caption{TDCCSD (full curves) energy-level populations computed with the reduced aug-cc-pVDZ
           basis set through CCLR projectors plotted as functions of time for CH$^+$.
           The dotted curves are populations computed with TDFCI theory.
           The total population of the black levels in the left panel is labelled
           ``Rest'' in the right panel. The curves labeled ``Sum'' are the total
           population of all $21$ levels computed at the CCSD and FCI levels of theory.}
  \label{fig:ch+_ravdz_pop}
\end{figure}

In order to make TDFCI simulations feasible for CH$^+$, we use a reduced aug-cc-pVDZ
basis set where the diffuse $p$ funtions on hydrogen and the diffuse $d$ functions
on carbon have been removed. While removing these diffuse functions has little effect
on the $5$ lowest CCSD linear response excitation energies (RMS deviation $0.001\,\au$),
the effect is significant on the following $25$ excitation energies with an RMS deviation
of $0.021\,\au$. The $31$ lowest-lying states, forming $21$ energy levels, computed with
CCSD linear response and FCI theory are shown in
the left panel of Fig.~\ref{fig:ch+_ravdz_pop}.
The ionization energy of CH$^+$ is estimated to be $0.878\,\au$ using ionization-potential
EOMCCSD theory and $0.876\,\au$ using FCI theory with the 6-31G* basis set.~\cite{Hirata2015}
Hence, all computed states are below the estimated ionization energies.

We expose the CH$^+$ ion to a sinusoidal laser pulse with intensity $2654\,\text{TW/cm}^2$
(${\cal E}_0 = 0.275\,\au$) and carrier frequency $\omega_0 = 0.212\,\au$,
which is resonant with the $^1\Pi \to\,^1\Sigma^+$ transition between excited states
at the CCSD level of theory.
The pulse is polarized along the $y$-axis, perpendicular to the bond axis,
with duration
$t_d = 66.7\,\au = 1.61\,\text{fs}$, corresponding to $2.05$ optical cycles.
The TDCCSD and TDFCI equations of motion were integrated with the 
sixth-order ($s=3$) Gauss-Legendre integrator with time step $\Delta t = 0.05\,\au$
and convergence threshold $10^{-6}$. The resulting energy-level populations
are presented in the right panel of Fig.~\ref{fig:ch+_ravdz_pop}.
Populations of the $4$ lowest-lying levels, including the ground state, are shown
along with the sum of the populations of the remaining $17$ levels, labelled ``Rest''.
The total population of all computed levels is labelled ``Sum''.

The effect of poorly described double-excited states at the CCSD level of theory is
evident, although we do observe a qualitative agreement with FCI theory.
We first note that the $21$ levels included in the analysis only account for
about $80\%$ of the norm of the FCI wave function at the end of the pulse,
implying that a physically correct description must also take ionization processes into account.
This, of course, is not surprising, considering the high intensity of the pulse.
The TDFCI and TDCCSD populations agree reasonably well during the first $0.75\,\text{fs}$ of
the simulation, wheras the TDCCSD errors increase as the simulation progresses.

\begin{figure}
  \includegraphics{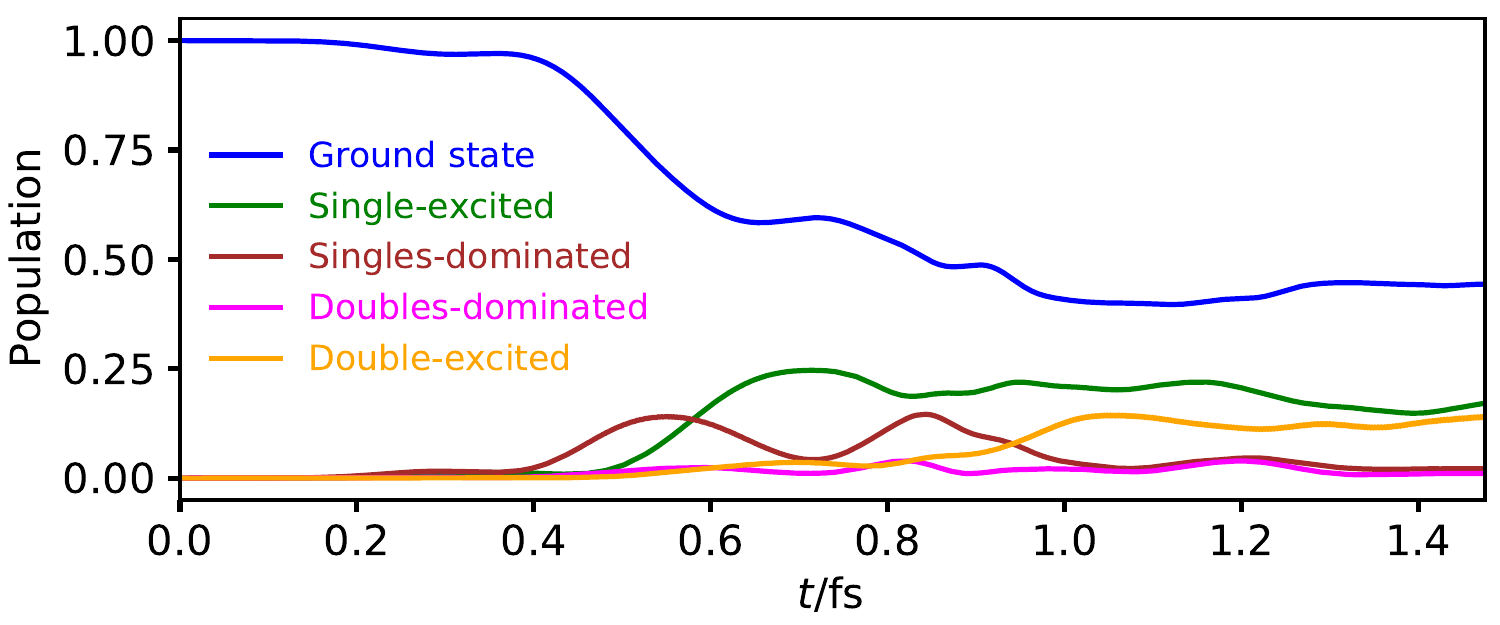}
  \caption{Population of different classes of CCSD states for CH$^+$ with the
           reduced aug-cc-pVDZ basis set.} 
  \label{fig:ch+_ravdz_total}
\end{figure}

Involving a large number of states, the dynamics is considerably more complex than the
dynamics of the cases presented above.
For simplicity, we use a classification of the $30$ excited
states based on their single- or double-excitation character.
Excited states with more than $90\%$ contribution from singles in the EOMCCSD
amplitude norm are classified as single-excited states, while states with less than 
$10\%$ singles contribution are classified as double-excited states. States with $10$--$90\%$
singles contribution are mixed states, classified as singles-dominated 
($>50\%$ singles contribution) or doubles-dominated ($<50\%$ singles contribution).
Thus, the $30$ excited CCSD states can be grouped into $7$ single-excited states,
$4$ singles-dominated states, $5$ doubles-dominated states, and $14$ double-excited states.
The total population of each class of states is presented in Fig.~\ref{fig:ch+_ravdz_total}.
After about $0.4\,\text{fs}$ the laser pulse induces transitions from the ground state
into singles-dominated states, followed by transitions (from both the ground state and the
singles-dominated excited states) into single-excitation states. Double-excited or
doubles-dominated states are barely populated at this stage, explaining the reasonable
agreement with TDFCI populations. Roughly half-way through the simulation, double-excited
states and, to a smaller extent, doubles-dominated states become populated, mainly
due to transitions from singles-dominated states. As soon as these processes occur, the
agreement between TDCCSD and TDFCI deteriorates.

\subsection{Population conservation}

As discussed in Section~\ref{subsec:conservation}, stationary-state populations in the absence of external forces
are strictly conserved in the FCI limit but may vary when the cluster operators are truncated. In order to investigate
the breaking of the population conservation law within TDCCSD theory, we have conducted simple numerical experiments
with several of the systems presented above. We apply the same
sinusoidal laser pulses as above but continue the propagation
after the pulses have been turned off, recording stationary-state population using the CCLR and EOMCC projectors.

\begin{figure}
  \includegraphics{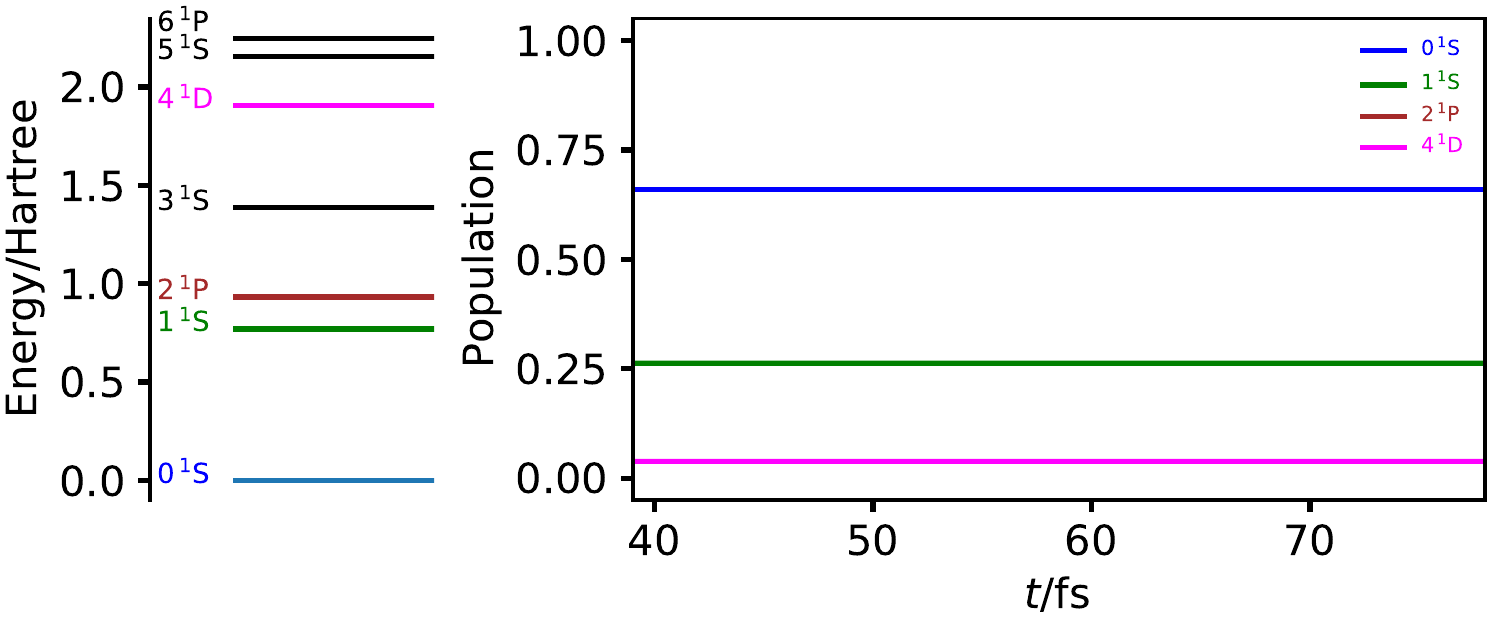}
  \caption{Conservation of TDCCSD energy-level populations after the laser pulses have been turned
           off for the He atom with the aug-cc-pVTZ basis set. The populations were computed using
           the CCLR projector.}
  \label{fig:he_avtz_pop_conservation}
\end{figure}

Figure \ref{fig:he_avtz_pop_conservation} shows the conservation of TDCCSD populations after the laser pulses
have been turned off for the He atom with the aug-cc-pVTZ basis set. 
A maximum absolute deviation in the populations of $1.5\times 10^{-3}$ is observed
for the $1^1\text{S}$ and $2^1\text{P}$ levels, whereas the deviations for the remaining levels
are at least one order of magnitude smaller. These deviations from exact conservation are likely caused
by the discretization of the numerical integration.

\begin{figure}
  \includegraphics{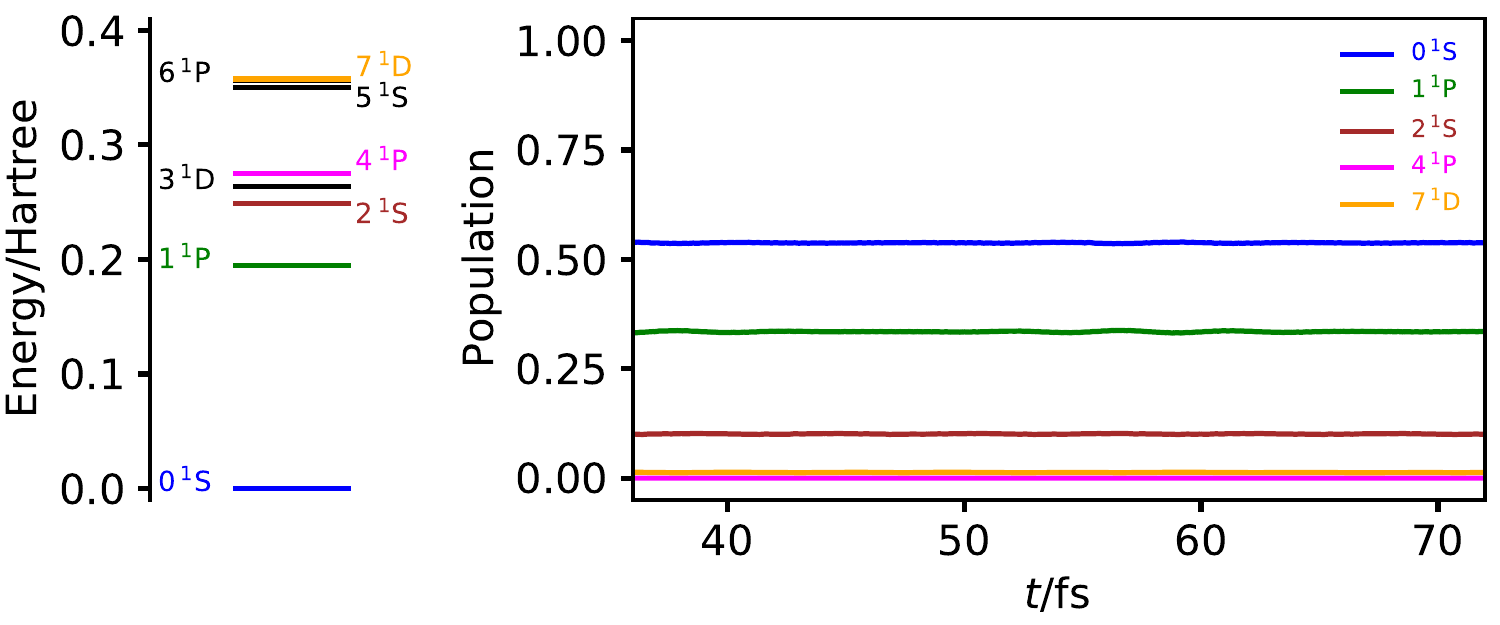}
  \caption{Conservation of TDCCSD energy-level populations after the laser pulses have been turned
           off for the Be atom with the aug-cc-pVTZ basis set. The populations were computed using
           the CCLR projector.}
  \label{fig:be_avtz_pop_conservation}
\end{figure}

Slightly larger deviations from strict conservation are observed for the Be atom with the aug-cc-pVTZ
basis set in Fig.~\ref{fig:be_avtz_pop_conservation}.
The maximum absolute deviation of $0.005$ is observed for the $1^1\text{P}$ level.
Caused by the truncation of the cluster operators in conjunction with discretization,
this deviation is thrice greater than that observed for He above.
Only a weak oscillatory behavior is observed, indicating that the states involved in the 
dynamics are very well approximated at the CCSD level of theory.

\begin{figure}
  \includegraphics{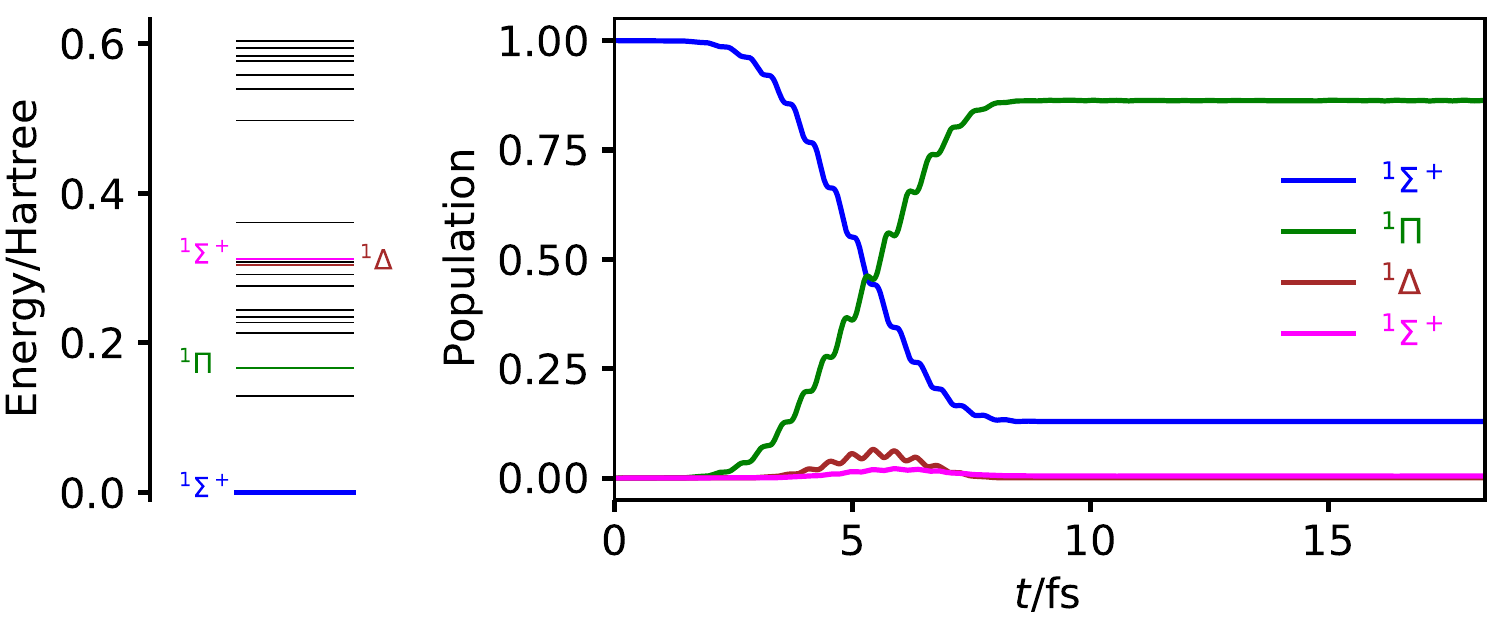}
  \caption{Conservation of TDCCSD energy-level populations of LiH after interaction with a chirped laser
           pulse with chirp rate $b=0.03078\,\text{fs}^{-2}$.
           The populations were computed using
           the CCLR projector and the aug-cc-pVDZ basis set.}
  \label{fig:lih_avdz_pop_conservation_bfac_0.9}
\end{figure}

Energy-level populations for LiH during and after interaction with a chirped laser pulse
are plotted Fig.~\ref{fig:lih_avdz_pop_conservation_bfac_0.9}.
The chirp rate $b=0.03078\,\text{fs}^{-2}$ yields a final state dominated by the $^1\Pi$ level and the populations remain
constant to an excellent approximation after the interaction ceases.
The maximum absolute deviation is observed for the $^1\Pi$ level and is on par with that observed for
He atom: $1.7 \times 10^{-3}$.

\begin{figure}
  \includegraphics{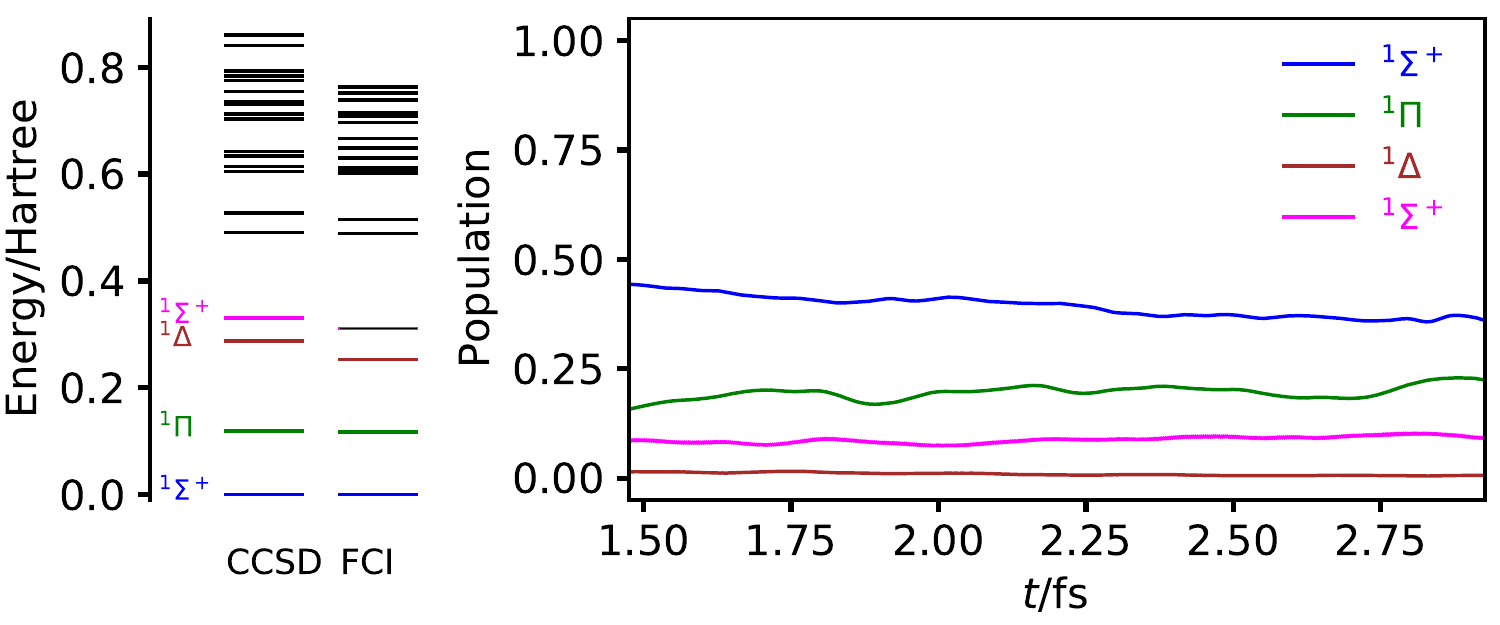}
  \caption{Conservation of TDCCSD energy-level populations of CH$^+$ after interaction with a laser pulse.
           The populations were computed using
           the CCLR projector and the reduced aug-cc-pVDZ basis set.}
  \label{fig:ch+_level_pop_ccfreq_conservation}
\end{figure}

As we saw above, the TDCCSD method is a much poorer approximation to TDFCI theory in the case of CH$^+$,
where double-excited states participate significantly in the dynamics. On these grounds, we expect a much poorer
conservation of energy-level populations after interaction with the laser pulse and, indeed, large
deviations can be seen in Fig.~\ref{fig:ch+_level_pop_ccfreq_conservation}.
The depletion of the ground state appears to continue after the pulse and irregular weakly oscillatory 
behavior is observed for the $^1\Pi$ level with a maximum absolute deviation of $0.072$, which is an order
magnitude greater than the devations found for He, Be, and LiH above.
The double-excited $^1\Delta$ and $^1\Sigma$ levels show maximum absolute deviations of $0.009$ and $0.019$,
respectively.

\subsection{Pump spectrum of LiF}

\citeauthor{Skeidsvoll2020} recently reported a theoretical TDCCSD study of
transient X-ray spectroscopy of the LiF molecule.~\cite{Skeidsvoll2020}
They applied a pump-probe laser setup with an optical pump pulse
resonant with the lowest-lying dipole-allowed transition from the ground state,
followed, at various delays,
by an X-ray probe pulse resonant with the first dipole-allowed core excitation.
The resulting time-resolved spectra were interpreted by means of excitation
energies from EOMCC theory and core-valence separated EOMCC theory.~\cite{Coriani2015}
The pump absorption spectrum reported in Fig. 7 of Ref.~\citenum{Skeidsvoll2020} contains
weak unassigned features, one weak absorption above the two low-lying valence excitations
and two very weak features below, which the authors speculated were due to two-photon
absorptions. We will now use the EOMCC and CCLR projectors to investigate what
might cause these weak features of the pump absorption spectrum.
We use the same basis set, denoted aug-cc-p(C)VDZ, as in Ref.~\citenum{Skeidsvoll2020}:
the aug-cc-pVDZ basis set for Li and the aug-cc-pCVDZ basis set for F.
The closed-shell TDCCSD equations of motion were integrated using
the sixth-order ($s=3$) Gauss-Legendre integrator
with time step $\Delta t = 0.025\,\au = 0.60\,\text{as}$ and convergence threshold $10^{-6}$ for
the fix-point iterations.

Initially in the ground state, we expose the LiF molecule to a shortened but otherwise identical
$z$-polarized Gaussian laser pulse to that in Ref.~\citenum{Skeidsvoll2020}, with
field strength ${\cal E}_0 = 0.01\,\au$ (peak intensity $I= 3.51\,\text{TW/cm}^2$),
carrier frequency $\omega_0 = 0.2536\,\au$, and
Gaussian RMS width $\sigma = 20\,\au$. The shortening consists in choosing
the central time $t_0 = 80\,\au$ (compared with $t_0 = 160\,\au$ in Ref.~\citenum{Skeidsvoll2020})
and $N=4$ (compared with $N=8$ in Ref.~\citenum{Skeidsvoll2020}). This implies that the
electric-field amplitude jumps from zero to $3.3\times 10^{-6}\,\au$ at $t=0\,\au$ and
from $3.3\times 10^{-6}\,\au$ to zero at $t=160\,\au$, whereas virtually no discontinuities
can be observed with the pulse parameters used
in Ref.~\citenum{Skeidsvoll2020}
(they are on the order of $10^{-16}\,\au$).
Since the pump pulse is quite weak,
the effects of these discontinuities on the populations are negligible, as can readily be verified 
using a simple eigenstate expansion analogous to the one used for He and Be in
Section \ref{subsec:excited_state_rabi_oscillations}.

\begin{figure}
  \includegraphics{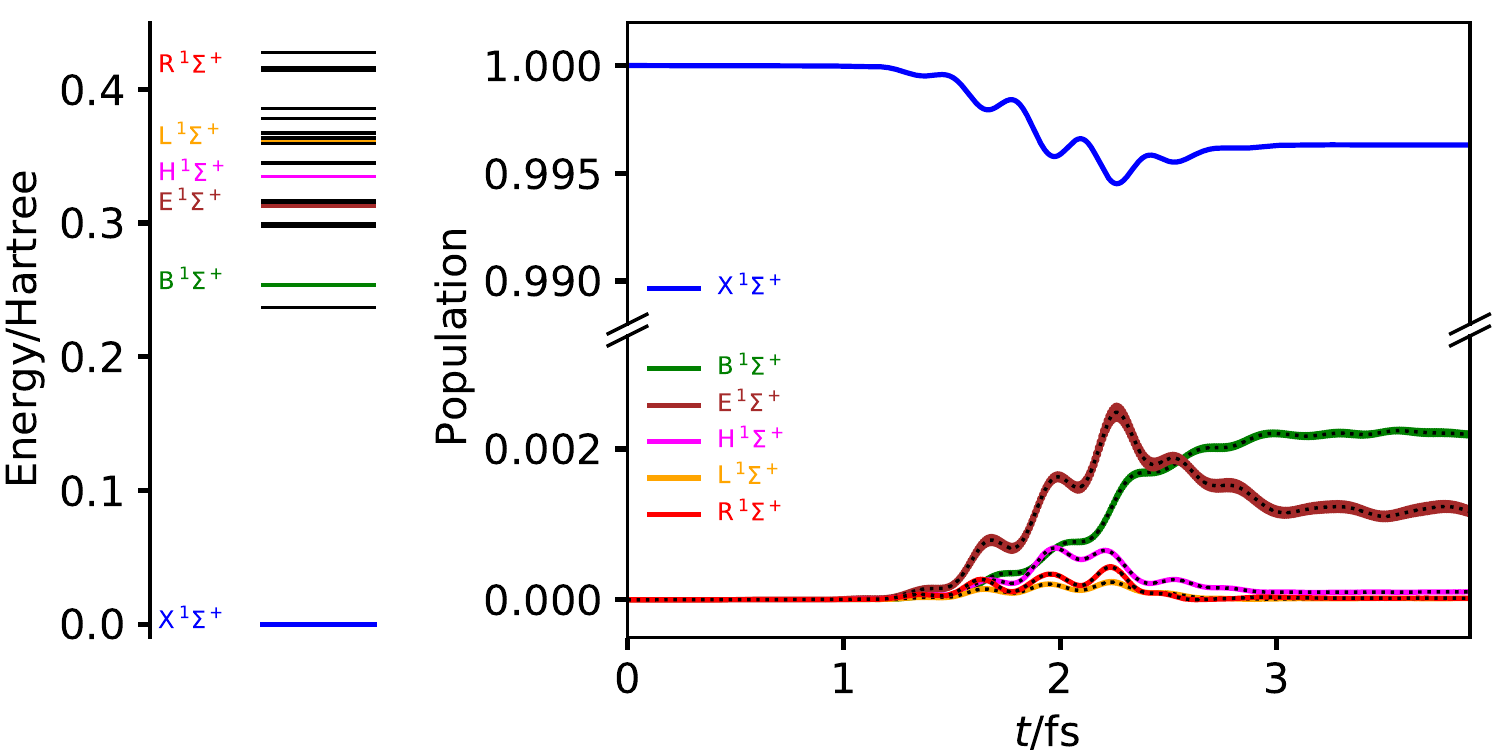}
  \caption{TDCCSD energy-level populations computed with the aug-cc-p(C)VDZ
           basis set through CCLR (colored full curves) and EOMCC (black dotted curves)
           projectors
           plotted as functions of time for LiF. The $20$ lowest-lying CCSD energy levels
           (corresponding to $30$ states)
           are shown in
           the left panel with participating levels counted alphabetically as in
           Ref.~\citenum{Skeidsvoll2020}.}
  \label{fig:lif_av_c_dz_pop}
\end{figure}

Our TDCCSD results are presented in Fig.~\ref{fig:lif_av_c_dz_pop}.
The first $30$ excited states ($20$ energy levels, left panel of Fig.~\ref{fig:lif_av_c_dz_pop})
are all single-excitation dominated states, with $94.9$--$95.4\%$ singles contribution to the
norm of the EOMCCSD amplitudes. The highest-lying states are somewhat above the first ionization
energy of LiF, which we estimate to be about $0.4\,\au$ ($11\,\text{eV}$) based on data available
in Ref.~\citenum{CCCBDB}.
The modest intensity of the pump pulse results in fairly little excitation from the ground state,
well within reach of a perturbation-theoretical (Fermi's golden rule) treatment.
In agreement with Ref.~\citenum{Skeidsvoll2020}, the projectors predict the absorption to
be dominated by the B$\,^1\Sigma^+$ and E$\,^1\Sigma^+$ states. The final population of
the latter is roughly $53\%$ of the former, in good agreement with the relative
intensities of the pump spectrum reported in Fig. 7 of Ref.~\citenum{Skeidsvoll2020}.
The population of the E$\,^1\Sigma^+$ state reaches its maximum value $0.0025$ at $t=2.26\,\text{fs}$;
at the same time the ground-state population reaches its minimum value $0.9945$, indicating that
the ensuing decay of the E$\,^1\Sigma^+$ population is caused by transition back to the
ground state.

The weak feature at higher frequency (at about $9.1\,\text{eV}$) in the pump spectrum of Ref.~\citenum{Skeidsvoll2020} is
seen to be consistent with one-photon transition from the ground state to the H$\,^1\Sigma^+$ state,
whose final population is about $5\%$ of that of the B$\,^1\Sigma^+$ state.
As speculated by \citeauthor{Skeidsvoll2020}~\cite{Skeidsvoll2020},
the two very weak features below the B$\,^1\Sigma^+$ line in Ref.~\citenum{Skeidsvoll2020}
are indeed seen to arise from direct two-photon absorptions from the ground state
to the L$\,^1\Sigma^+$ and R$\,^1\Sigma^+$ states. The only alternative explanation would be excitations
between excited states, but this mechanism can almost certainly be ruled out, since the population of 
these states starts before other excited levels are significantly populated and since
no other excited states are depleted as the populations of these states increase.

Although the CCLR and EOMCC populations largely agree with an overall RMS deviation
of $7 \times 10^{-6}$, the CCLR populations show 
spurious high-frequency oscillations in Fig.~\ref{fig:lif_av_c_dz_pop}.
The oscillations are caused by the off-diagonal contributions from
$\bra{\breve{\Psi}_n}$ (eq.~\eqref{eq:breve_psi_n}) to the CCLR projector (eq.\eqref{eq:cclr_projector}).
While these contributions are required to ensure proper size-intensivity
of one-photon transition moments, they also cause nonorthogonality of the CCLR excited-state representation
as expressed by eq.~\eqref{eq:cclr_orth}.
Since the CCLR and EOMCC projectors provided virtually identical results in the cases above,
this observation serves as a recommendation
of the EOMCC projector for the calculation of stationary-state populations.

\begin{figure}
  \includegraphics{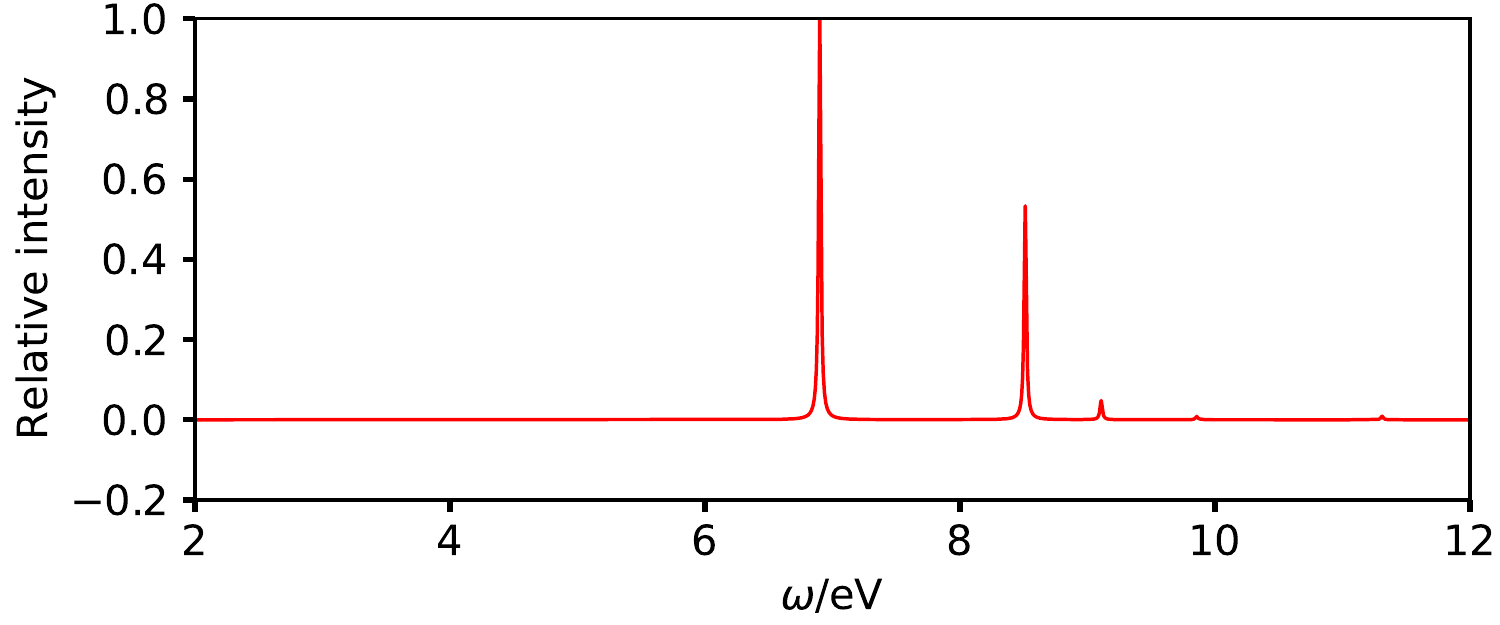}
  \caption{Pump spectrum of LiF generated from EOMCC populations assuming one-photon
           transitions from the ground state.}
  \label{fig:lif_spectrum}
\end{figure}

Figure~\ref{fig:lif_spectrum} depicts a normalized pump spectrum of LiF
generated from the final EOMCC populations using
\begin{equation}
   I(\omega) = \frac{S(\omega)}{\max_\omega S(\omega)}, \qquad
   S(\omega) = \sum_n \frac{p_n(t_\text{\scriptsize{final}})}{\pi}
               \frac{\gamma}{(\omega - \Delta E_n)^2 + \gamma^2}
\end{equation}
where $\gamma = 0.01\,\text{eV}$ is an artifical Lorentzian broadening of the excited levels.
This approach implicitly assumes that the excited states only become populated
through one-photon absorption from the ground state, thus excluding
all nonlinear optical processes.
The population-based pump spectrum agrees remarkably well with that reported by
\citeauthor{Skeidsvoll2020}~\cite{Skeidsvoll2020}, which was properly generated from
Fourier transformation of the induced dipole moment.
This supports the conclusion that the low-frequency features are two-photon absorptions
and further strengthens the confidence in the proposed EOMCC projector for the calculation
of stationary-state populations in TDCC simulations.

\section{Concluding remarks}
\label{sec:concluding_remarks}

We have proposed projectors for the interpretation of many-electron dynamics
within TDCC theory in terms of the population of stationary states.
Two conditions are used to define suitable projectors from CC linear response
theory and from EOMCC theory: (\emph{i}) the projector must reproduce the 
correct form of one-photon transition strengths and (\emph{ii}) the projectors
must yield populations that converge to the FCI results in the limit of
untruncated cluster operators.

The CCLR and EOMCC projectors are tested numerically at the TDCCSD level of theory
for the laser-driven dynamics in the He and Be atoms, and in
the molecules LiH, CH$^+$, and LiF. It is demonstrated that the populations
provide valuable insight into the linear and nonlinear optical
processes occcuring during the interaction of the electrons with laser pulses.
For the He atom, it is verified numerically
that the populations computed from both CCLR and EOMCC projectors agree
with those computed from TDFCI simulations. For Be and LiH, the CCLR and
EOMCC populations show excellent agreement with TDFCI populations,
since the excited stationary states involved in the dynamics are dominated by
single-excited Slater determinants. Such states are generally well described by
CCSD theory. For CH$^+$, we deliberately design the laser pulse such that
double-excitation dominated states become populated, which 
reduces the agreement between TDFCI and TDCCSD populations. 
This is also reflected in studies of the conservation of populations after
the laser pulses are turned off. Theoretically, the TDCC populations will only be
strictly conserved in the FCI limit. Numerically, we find that they are nearly
conserved as long as the participating stationary states are well approximated
at the CCSD level of theory.
Finally, for LiF, we use the CCLR and EOMCC projectors to explain unassigned
weak features in a theoretical TDCCSD pump spectrum reported recently~\cite{Skeidsvoll2020}.

Overall, the CCLR and EOMCC projectors yield very similar excited-state 
populations with typical
RMS deviations on the order of $10^{-5}$. For LiF, however, we observe small-amplitude
high-frequency oscillations of the excited-state populations computed with the
CCLR projector. Originating from a contribution that vanishes in the
FCI limit, we speculate that such oscillations may increase for larger and more
complex systems where TDCCSD theory may be further from TDFCI theory.
Not showing signs of such spurious behavior, the EOMCC projector
appears more attractive than the CCLR projector.
This has the added benefit that the additional response equations \eqref{eq:Mbar_eqs}
need not be solved, thus making the EOMCC projectors less computationally
demanding than the CCLR projectors.

These findings call for further research aimed at a fully consistent definition of
excited states in CC theory, and work in this direction is in progress in our labs.

\begin{acknowledgement}

This work was supported by the Research Council of Norway (RCN) through its Centres of Excellence scheme, project number 262695,
by the RCN Research Grant No. 240698,
by the European Research Council under the European Union Seventh
Framework Program through the Starting Grant BIVAQUM, ERC-STG-2014 grant
agreement No. 639508, and by
the Norwegian Supercomputing Program (NOTUR) through a grant of computer time (Grant No.\ NN4654K).

\end{acknowledgement}

%
%

\appendix
\section*{Appendix: the FCI limit}

The CCLR projector, eq.~\eqref{eq:cclr_projector} becomes identical to 
the EOMCC projector, eq.~\eqref{eq:conventional_eom_projector}, in the
FCI limit. To show this, we
will now demonstrate that
\begin{equation}
  \bra{\breve{\Psi}_n} = \bra{\bar{\Psi}_n} - {\cal R}^n_0\bra{\tilde{\Psi}_0} = 0
\end{equation}
in the FCI limit.

In the FCI limit, the EOMCC wave functions satisfy the 
time-independent Schr{\"o}dinger equation and its hermitian
conjugate,
\begin{alignat}{2}
\label{eq:tise_cc_r}
   &H_0 \ket{\Psi_0} = \ket{\Psi_0} E_0, &\qquad
   &H_0 \ket{\Psi_n} = \ket{\Psi_n} E_n
\\
\label{eq:tise_cc_l}
   &\bra{\tilde{\Psi}_0} H_0 = E_0 \bra{\tilde{\Psi}_0}, &\qquad
   &\bra{\tilde{\Psi}_n} H_0 = E_n \bra{\tilde{\Psi}_n}
\end{alignat}
where the ground- and excited-state wave functions constitute a biorthonormal
set according to eqs.~\eqref{eq:eom_orth} and \eqref{eq:eom_orth_gs}.
The resolution-of-the-identity reads
\begin{equation}
\label{eq:cc_ri}
   \ket{\Psi_0}\bra{\tilde{\Psi}_0}
   +
   \sum_n \ket{\Psi_n}\bra{\tilde{\Psi}_n}
   = 1
\end{equation}
where $1$ is to be understood as the identity operator.
To verify the resolution-of-the-identity, one simply inserts the definitions
of the wave functions and exploits the biorthonormality of the Jacobian
eigenvectors, eq.~\eqref{eq:eigen},
along with the completeness of the underlying determinant basis.

According to eq.~\eqref{eq:Mbar_eqs}, the amplitudes $\bar{\cal M}^n_\mu$
can be recast as
\begin{align}
   \bar{\cal M}^n_\mu
   &=
   - \sum_{\nu\gamma} {\cal R}^n_\nu F_{\nu\gamma}
     \left(\boldsymbol{A} + \Delta E_n \boldsymbol{1}\right)^{-1}_{\gamma\mu}
\nonumber \\
   &=
   - \sum_m \sum_{\nu\gamma} {\cal R}^n_\nu F_{\nu\gamma} 
     {\cal R}^m_\gamma (\Delta E_n + \Delta E_m)^{-1} {\cal L}^m_\mu
\nonumber \\
   &=
   - \sum_m \braket{\tilde{\Psi}_0 \vert [[H_0,R_n],R_m] \vert \Psi_0}
     (\Delta E_n + \Delta E_m)^{-1} {\cal L}^m_\mu
\end{align}
where we have used eq.~\eqref{eq:eigen}.
Expanding the nested commutator and
using eqs.~\eqref{eq:tise_cc_r}, \eqref{eq:tise_cc_l}, \eqref{eq:cc_ri},
and \eqref{eq:eom_orth} we find
\begin{align}
   \braket{\tilde{\Psi}_0 \vert [[H_0,R_n],R_m] \vert \Psi_0}
   &=
   -(\Delta E_n + \Delta E_m)
   \left(
     \braket{\tilde{\Psi}_0 \vert R_nR_m\vert\Psi_0} - {\cal R}^n_0{\cal R}^m_0
   \right)
\nonumber \\
   &=
   -(\Delta E_n + \Delta E_m)
    \braket{\tilde{\Psi}_0 \vert R_n \left( 1 - \ket{\Psi_0}\bra{\tilde{\Psi}_0}\right)
            R_m \vert \Psi_0}
\nonumber \\
   &=
   -(\Delta E_n + \Delta E_m)
   \braket{\tilde{\Psi}_0 \vert R_n \vert \Psi_m}
\end{align}
giving
\begin{equation}
   \bar{\cal M}^n_\mu = \sum_m \braket{\tilde{\Psi}_0 \vert R_n \vert \Psi_m} {\cal L}^m_\mu
\end{equation}
Hence,
\begin{align}
   \bra{\bar{\Psi}_n}
   &=
   \sum_\mu \bra{\tilde{\Phi}_\mu} \bar{\cal M}^n_\mu \ee^{-T_0} - \bra{\tilde{\Psi}_0} R_n
\nonumber \\
   &=
   \sum_m \braket{\tilde{\Psi}_0 \vert R_n \vert \Psi_m} \bra{\tilde{\Psi}_m}
   - \bra{\tilde{\Psi}_0} R_n
\nonumber \\
   &=
   \bra{\tilde{\Psi}_0} R_n
   \left(
      \sum_m \ket{\Psi_m}\bra{\tilde{\Psi}_m} - 1
   \right)
\nonumber \\
   &=
   -\braket{\tilde{\Psi}_0 \vert R_n \vert \Psi_0}\bra{\tilde{\Psi}_0}
   = {\cal R}^n_0 \bra{\tilde{\Psi}_0}
\end{align}
which shows that $\bra{\breve{\Psi}_n} = 0$ in the FCI limit.
\bibliography{manuscript}

\end{document}